\documentclass[longauth]{aa}  

\usepackage{graphicx}
\usepackage{txfonts}

\usepackage{natbib}
\bibpunct{(}{)}{;}{a}{}{,} 		

\usepackage{enumitem}
\usepackage{lscape}

\usepackage{longtable}

\newcommand\T{\rule{0pt}{2.8ex}}       	
\newcommand\B{\rule[-1.8ex]{0pt}{0pt}} 	

\newcommand{\hto}{{\hbox {H\textsubscript{2}O}}}

\mathchardef\mhyphen="2D

\begin{document}

\title{$z$-GAL -- A NOEMA spectroscopic redshift survey of bright {\it Herschel} galaxies: [III] Physical properties}

\author{S.~Berta \inst{1}		\and
        F.~Stanley\inst{1}		\and 		
        D.~Ismail\inst{2}		\and 		
        P.~Cox\inst{3}              	\and   		
        R.~Neri \inst{1}            	\and 		
        C.~Yang\inst{4}            	\and 		
        A.~J. Young\inst{5}         	\and  		
        S.~Jin\inst{6,7}            	\and  		
        H.~Dannerbauer\inst{8,9}   	\and  		
        T.~J.~L.~C.~Bakx\inst{4,10,11}	\and 		
        A.~Beelen \inst{2}         	\and       	
        A.~Wei{\ss}\inst{12}         	\and		
        A.~Nanni\inst{13,14}		\and 		
        A.~Omont\inst{3}            	\and  		
        P.~van der Werf\inst{15}    	\and  		
        M.~Krips\inst{1}            	\and  		
        A.~J. Baker\inst{5,16}         	\and 		
        G.~Bendo\inst{17}		\and 		
        E.~Borsato\inst{18}		\and 		
        V.~Buat\inst{2}            	\and		
        K.~M. Butler\inst{1}          	\and		
        N.~Chartab\inst{19}		\and		
        A.~Cooray\inst{19}          	\and  		
        S.~Dye\inst{20}             	\and		
        S.~Eales\inst{21}            	\and		
        R.~Gavazzi\inst{3}          	\and 		
        D.~Hughes\inst{22}          	\and 		
        R.~J.~Ivison\inst{23,24,25,26}	\and 		
        B.~M.~Jones\inst{27}		\and		
        M.~Lehnert\inst{28}          	\and		
        L.~Marchetti\inst{29,30}   	\and		
        H.~Messias\inst{31,32}        	\and 		
        M.~Negrello\inst{21}         	\and 		
        I.~Perez-Fournon\inst{8,9}  	\and 		
        D.~A.~Riechers\inst{27}  	\and 		
        S.~Serjeant\inst{33}        	\and  		
        S.~Urquhart\inst{33}        	\and		
        C.~Vlahakis\inst{34}        	 		
}

\institute{Institut de Radioastronomie Millim\'etrique (IRAM), 300 rue de la Piscine, 38400 Saint-Martin-d'H{\`e}res, France\\
              \email{berta@iram.fr}
        \and 
             Aix-Marseille Universit\'{e}, CNRS and CNES, Laboratoire d'Astrophysique de Marseille, 38 rue Frédéric Joliot-Curie, 13388 Marseille, France
        \and
             Sorbonne Universit{\'e}, UPMC Universit{\'e} Paris 6 and CNRS, UMR 7095, Institut d'Astrophysique de Paris, 98b boulevard Arago, 75014 Paris, France
        \and
             Department of Space, Earth and Environment, Chalmers University of Technology, Onsala Space Observatory, SE-439 92 Onsala, Sweden
        \and  
             Department of Physics and Astronomy, Rutgers, The State University of New Jersey, 136 Frelinghuysen Road, Piscataway, NJ 08854-8019, USA
	\and
             Cosmic Dawn Center (DAWN), Radmandsgade 62, 2200 Copenhagen N, Denmark
        \and
             DTU Space, Technical University of Denmark, Elektrovej 327, DK-2800 Kgs. Lyngby, Denmark
        \and
             Instituto Astrof{\'i}sica de Canarias (IAC), E-38205 La Laguna, Tenerife, Spain  
        \and 
             Universidad de La Laguna, Dpto. Astrof{\'i}sica, E-38206 La Laguna, Tenerife, Spain
        \and
             Division of Particle and Astrophysical Science, Graduate School of Science, Nagoya University, Aichi 464-8602, Japan
        \and  
             National Astronomical Observatory of Japan, 2-21-1, Osawa, Mitaka, Tokyo 181-8588, Japan
        \and  
             Max-Planck-Institut f{\"u}r Radioastronomie, Auf dem H{\"u}gel 69, 53121 Bonn, Germany.
	\and
	     National Centre for Nuclear Research, ul. Pasteura 7, 02-093 Warsaw, Poland
	\and
	     INAF -- Osservatorio astronomico d'Abruzzo Via Maggini SNC 64100 Teramo
        \and 
             Leiden University, Leiden Observatory, PO Box 9513, 2300 RA Leiden, The Netherlands
        \and
	     Department of Physics and Astronomy, University of the Western Cape, Robert Sobukwe Road, Bellville 7535, South Africa
	\and
	     UK ALMA Regional Centre Node, Jodrell Bank Centre for Astrophysics, Department of Physics and Astronomy, The University of Manchester, Oxford Road, Manchester M13 9PL, United Kingdom
	\and
	     Dipartimento di Fisica e Astronomia ``G. Galilei'', Universit{\`a} di Padova, vicolo dell’Osservatorio 3, I-35122 Padova, Italy
	\and
             University of California Irvine, Department of Physics \& Astronomy, FRH 2174, Irvine CA 92697, USA
	\and 
             School of Physics and Astronomy, University of Nottingham, University Park, Nottingham NG7 2RD, UK
        \and
             School of Physics and Astronomy, Cardiff University, Queens Building, The Parade, Cardiff CF24 3AA, UK
        \and
             Instituto Nacional de Astrofísica, {\'O}ptica y Electr{\'o}nica, Astrophysics Department, Apdo 51 y 216, Tonantzintla, Puebla 72000 Mexico
        \and
             European Southern Observatory, Karl-Schwarzschild-Strasse 2, 85748 Garching, Germany
        \and
             Department of Physics and Astronomy, Macquarie University, North Ryde, New South Wales, Australia
	\and
             School of Cosmic Physics, Dublin Institute for Advanced Studies, 31 Fitzwilliam Place, Dublin D02 XF86, Ireland
	\and
             Institute for Astronomy, University of Edinburgh, Royal Observatory, Blackford Hill, Edinburgh EH9 3HJ
	\and
             I. Physikalisches Institut, Universit{\"a}t zu K{\"o}ln, Z{\"u}lpicher Strasse 77, D-50937 K{\"o}ln, Germany
	\and 
	     Centre de Recherche Astrophysique de Lyon - CRAL, CNRS UMR 5574, UCBL1, ENS Lyon, 9 avenue Charles Andr\'e, F-69230 Saint-Genis-Laval, France
	\and
             University of Cape Town, Department of Astronomy. Private Bag X3 Rondebosch, 7701 Cape Town, South Africa
        \and 
             INAF - Instituto di Radioastronomia - Italian ARC, Via Piero Gobetti 101, 40129, Bologna, Italy
        \and  
             Joint ALMA Observatory, Alonso de C{\'o}rdova 3107, Vitacura 763-0355, Santiago de Chile, Chile
        \and  
             European Southern Observatory, Alonso de C{\'o}rdova 3107, Casilla 19001, Vitacura, Santiago, Chile
	 \and
             Department of Physical Sciences, The Open University, Milton Keynes MK7 6AA, UK
        \and
             National Radio Astronomy Observatory, 520 Edgemont Road, Charlottesville VA 22903, USA
    }

\date{Received: 3 May 2023 / accepted: 14 July 2023}


\abstract{
The $z$-GAL survey observed 137 bright {\it Herschel}-selected targets with the IRAM Northern Extended Millimeter Array, with the aim to measure their redshift and study their properties. Several of them have been resolved into multiple sources. Consequently, robust spectroscopic redshifts have been measured for 165 individual galaxies in the range $0.8<z<6.5$. In this paper we analyse the millimetre spectra of the $z$-GAL sources, using both their continuum and line emission to derive their physical properties.
At least two spectral lines are detected for each source, including transitions of $\rm ^{12}CO$, [C{\small I}], and \hto. 
The observed $\rm ^{12}CO$ line ratios and spectral line energy distributions of individual sources resemble those of local starbursts. In seven sources the para-$\rm H_2O(2_{11}-2_{02})$ transition is detected and follows the IR versus \hto\ luminosity relation of sub-millimetre galaxies.
The molecular gas mass of the $z$-GAL sources is derived from their $\rm ^{12}CO$, [C{\small I}], and sub-millimetre dust continuum emission. The three tracers lead to consistent results, with the dust continuum showing the largest scatter when compared to $\rm ^{12}CO$. 
The gas-to-dust mass ratio of these sources was computed by combining the information derived from $\rm ^{12}CO$ and the dust continuum and has a median value of 107, similar to star-forming galaxies of near-solar metallicity.
The same combined analysis leads to depletion timescales in the range between 0.1 and 1.0 Gyr, which place the $z$-GAL sources between the `main sequence' of star formation and the locus of starbursts.
Finally, we derived a first estimate of stellar masses -- modulo possible gravitational magnification -- by inverting known gas scaling relations: the $z$-GAL sample is confirmed to be mostly composed by starbursts, whereas $\sim25$\% of its members lie on the main sequence of star-forming galaxies (within $\pm0.5$ dex).
}

\keywords{Submillimeter: galaxies -- Galaxies: high-redshift -- Galaxies: starburst -- Galaxies: star formation -- Galaxies: statistics -- Galaxies: ISM}

\authorrunning{S. Berta et al.}

\titlerunning{$z$-GAL III -- physical properties}

\maketitle


\section{Introduction}\label{sect:intro}

Galaxy star formation takes place in dense gas clouds, fuelled by molecular hydrogen and catalysed by dust.  
Enriched gas is expelled in the form of stellar and galactic winds, to then be partially recycled to form new stars \citep[e.g.][]{mckee2007,bouche2010,kennicutt2012,lilly2013,tacconi2020}.

The discovery and identification by the Infrared Space Observatory (ISO), the {\it Spitzer} Space Telescope, and the {\it Herschel} satellite of large numbers of distant sources emitting a substantial amount of their energy in the infrared \citep[IR, e.g.][]{smail1997, aussel1999, elbaz1999, lonsdale2003, papovich2004, frayer2009, eales2010, lutz2011, elbaz2011, oliver2012} demonstrated that, although locally rare, powerful IR galaxies are numerous at high redshift. 
The UV-optical emission of the newly formed young stars is absorbed by dust and re-processed into the far-infrared (far-IR) as thermal emission. 

The majority of the star formation at high redshift occurs in dusty star-forming galaxies \citep[DSFGs, e.g.][]{bourne2017, bouwens2016, bouwens2020, dunlop2017, hatsukade2018, dudzevicciute2020, zavala2021}. 
The peak of galaxy growth, traced by the cosmic star formation density (SFRD) occurred at redshifts $1<z<3$ \citep[see the review by][]{madaudickinson2014}. {\it Spitzer} and {\it Herschel} extragalactic surveys demonstrated that in the local Universe the cosmic SFRD is dominated by galaxies with IR luminosity $L_\textrm{IR}<10^{11}$ L$_\odot$, while luminous infrared galaxies (LIRGs with $L_\textrm{IR}>10^{11}$ L$_\odot$) dominate at redshift $z>1$ and ultra-luminous infrared galaxies (ULIRGs, $L_\textrm{IR}>10^{12}$ L$_\odot$) at $z>2$ \citep[e.g.][]{magnelli2011,magnelli2013}.

Star-forming galaxies occupy a preferential locus in the stellar mass versus star formation rate (SFR) space, called the star formation `main sequence' \citep[MS, e.g.][]{guzman1997, brinchmann2000, papovich2006, noeske2007, elbaz2007, daddi2007}. The MS exists at all redshifts and its normalisation -- that is its SFR for a given stellar mass -- increases at earlier cosmic times \citep[e.g.][]{elbaz2007}. 
Galaxies lying above the MS experience intense events of star formation: starbursts possibly triggered by galaxy interactions and mergers that are destined to exhaust their gas reservoir over timescales much shorter than the Hubble time. On the other hand, MS galaxies undergo a `secular evolution' characterised by a regular and constant star formation activity \citep[e.g.][]{saintonge2013, tacconi2018, tacconi2020}. 

Given the evolution of the MS as a function of redshift, a galaxy with $\textrm{SFR}\sim100$ M$_\odot$/yr is a powerful starburst in the local Universe, but is a secularly evolving MS galaxy at $z>2$ \cite[e.g.][]{elbaz2010,nordon2010}. Consequently, galaxy growth and the cosmic SFRD are dominated by MS galaxies through time and at least up to  $z\sim3$.
Starbursts, on the other hand, contribute only to a fraction 5-10\% of the cosmic SFRD \citep[e.g.][]{rodighiero2011, sargent2014, schreiber2015}.

The inferred IR luminosity of DSFGs can reach 10$^{13}$\,L$_\odot$ \citep[see reviews by][]{blain2002,casey2014,hodge2020}, corresponding to a $\textrm{SFR}\sim 1000$ M$_\odot$ yr$^{-1}$. 
To sustain such a large SFR, a significant reservoir of molecular gas is required. Quantifying the molecular gas reservoir of DSFGs is imperative in order to understand the star formation processes they are undergoing and their evolution. 

As the direct measurement of the $\textrm{H}_2$ mass is hindered by the difficulty to directly detect the $\textrm{H}_2$ molecule in cold molecular clouds (Sect. \ref{sect:mgas}), other tracers associated with $\textrm{H}_2$ have been adopted. The most used tracer is carbon monoxide ($\rm ^{12}CO$), the second most abundant component of the molecular reservoir of galaxies after $\textrm{H}_2$, and bright enough to allow for observations at high-$z$. With a good sampling of the $\rm ^{12}CO$ spectral line energy distribution (SLED), it is possible to derive the physical properties of the molecular interstellar medium (ISM) such as its density, kinetic temperature, and mass \citep[see reviews by e.g.][]{Carilli-Walter2013,combes2018}. 
In addition to $\rm ^{12}CO$, the \hto\ and [C{\small I}] transitions have also proved to be valuable tracers of the molecular gas reservoir in high-redshift galaxies \citep[e.g.][and references therein]{Omont2013, Yang2020, Yang2016, valentino2020, dunne2022}, while other molecules tracing dense gas such as HCN or HCO$^+$ are too faint to be observed in large samples of high-$z$ sources \citep[e.g.][]{rybak2022}. 
Spectroscopic observations of $\rm ^{12}CO$ at $z>1$ are to date still limited to a few hundreds of galaxies and to a much smaller number for [C{\small I}] and \hto. A promising alternative is the dust sub-millimetre continuum of galaxies, which is much less expensive than spectroscopy in terms of observing time \citep[e.g.][]{scoville2014,scoville2016,scoville2023}. 

The star formation of a galaxy is related to its fuel by the so-called Kennicutt-Schmidt law \citep{schmidt1959, kennicutt1998b}, which links its SFR to its molecular gas mass (or density). The ratio between these two quantities represents the timescale over which the galaxy would deplete its entire fuel reservoir at the current rate of star formation. 
Starbursts would consume all their molecular gas over timescales of few $10^8$ years or shorter \citep{genzel2010, genzel2015, tacconi2018, tacconi2020}. 

Despite their modest contribution to the global star formation and galaxy assembly budget, starburst are still the sites where the most extreme activity and physical processes take place. 
The most luminous DSFGs -- for example, selected with bright flux cuts at far-IR millimetre observed wavelengths -- have been and are the subject of extensive multi-wavelength follow-up observations with the aim to constrain their redshift and measure their molecular gas properties \citep[e.g.][]{weiss2009, walter2012, harris2012, lupu2012, weiss2013, strandet2016, fudamoto2017, danielson2017, reuter2020, urquhart2022}.

{\it Herschel} sources selected to have 500 $\mu$m fluxes above 80-100 mJy are found over a very wide redshift range \citep[$1<z<6$]{nayyeri2016, bakx2018, neri2020, reuter2020, urquhart2022}. 
A large number of them are gravitationally lensed  \citep[e.g.][]{negrello2010, negrello2017, conley2011, riechers2011d, cox2011, wardlow2013, bussmann2013, nayyeri2016, bakx2020b, bakx2020}, while others have been resolved into galaxy groups \citep[e.g.][]{bussmann2015, oteo2018, gomez-guijarro2019, ivison2019}. In rare cases intrinsically hyper-luminous infrared galaxies (HyLIRGs, with $L_{\rm IR} > 10^{13}$ L$_\odot$) have been identified \citep[e.g.][]{ivison2013, ivison2019, fu2013, oteo2016, riechers2013, riechers2017}.
Targeted CO observations have revealed large molecular gas masses of 10$^{10-11}$ M$_\odot$ \citep[e.g.][]{tacconi2008, ivison2011, bothwell2013, aravena2016, harrington2021, stanley2022}, but have a diversity of excitation properties, reflected in their different SLEDs \citep[e.g.][]{yang2017, stanley2022}.

The $z$-GAL survey, carried out with the IRAM Northern Extended Millimeter Array (NOEMA), is designed to study a large sample of {\it Herschel}-selected IR-bright DSFGs to determine their redshift, measure their dust and molecular gas content, and study their detailed nature. Following the success of a pilot project, which reported reliable spectroscopic redshifts of 11 such sources \citep{neri2020}, further 126 DSFGs with 500 $\mu$m fluxes brighter than 80 mJy were targeted. Thanks to the detection of multiple CO lines and occasional other species ($\rm H_2O$, [C{\small I}], and HCN/HCO$^+$), the success rate of $z$-GAL is 98.5\%, with robust spectroscopic redshifts measured for 135 out of the original 137 {\it Herschel}-selected targets \citep{cox2022}.

In \cite{stanley2022} a detailed analysis of the individual molecular gas properties of the $z$-GAL pilot sources was presented. By combining the NOEMA CO measurements from \cite{neri2020} with Very Large Array (VLA) follow-up observations targeting the $\rm ^{12}CO(1-0)$ line, a diversity in properties was revealed. The depletion times measured were consistent with galaxies being both on the main sequence and the starburst phase, demonstrating that even with a selection of the most luminous sources, we are still probing galaxies on the main sequence. Furthermore, even with a relatively small sample, a large variety in SLEDs and line ratios was found, covering the full range of what has been previously observed for DSFGs. With a sample that is more than ten times larger, $z$-GAL offers the opportunity to explore this diversity in more detail. 

In this paper, the third in the $z$-GAL series, we present the properties of the full $z$-GAL sample including the pilot project, as inferred combining their spectral lines and continuum emission. Paper I \citep{cox2022} presents the survey and an overview of the main results. The dust properties of the $z$-GAL sources are presented in Paper II \citep{ismail2022}. The last paper in this series, Paper IV (Bakx et al., in prep.) will discuss the lensed nature of the $z$-GAL sources.

This paper is organised as follows. Section \ref{sect:obs} briefly recalls the basic information about the $z$-GAL survey and describes the available data. Section \ref{sect:properties_mol_gas} presents the main properties of the $\rm ^{12}CO$ spectral lines, their luminosities, their ratios, and their energy distribution. The properties of water lines are studied in Sect. \ref{sect:H2O_line}. The molecular gas mass of the $z$-GAL sources, as derived from $\rm ^{12}CO$, [C{\small I}] and the sub-millimetre dust continuum, are computed and compared in Sect. \ref{sect:mgas}. Section \ref{sect:delta_GDR} reports on the gas-to-dust ratio of the targets. The integrated Kennicutt-Schmidt relation is studied in Sect. \ref{sect:KS}, together with molecular gas depletion timescales. Finally in Sect. \ref{sect:mstar} we perform an inversion of the depletion timescales scaling relation and derive a first estimate of the stellar mass of our sources. Section \ref{sect:conclusions} summarises the main findings of this study.

Throughout this paper we adopt a spatially flat $\Lambda$CDM cosmology 
with $H_{0}=67.4\,{\rm km\,s^{-1}\,Mpc^{-1}}$ and $\Omega_\mathrm{M}=0.315$ \citep{planck2020_2018_VI} and we assume a \citet{chabrier2003} initial mass function (IMF). Several $z$-GAL sources are potentially amplified by gravitational lensing \citep[e.g.][]{berta2021}. The unknown magnification factor $\mu$ is therefore hereby explicitly written for all affected quantities.


\section{Survey and data overview}\label{sect:obs}
 
The $z$-GAL NOEMA Large Programme (project IDs M18AB and D20AB, PIs P.~Cox, H.~Dannerbauer, T.~Bakx) and Pilot Programme \citep[project IDs W17DM and S18CR, PI A.~Omont;][]{neri2020} observed a total of 137 far-IR-bright distant targets. These sources were selected from the {\it Herschel} Bright Sources \citep[HerBS][]{bakx2018}, the HerMES Large Mode Survey (HeLMS) and the {\it Herschel} Stripe 82 (HeRS) Survey \citep{nayyeri2016}. 

The HeLMS and HerS surveys include {\it Herschel} sources with $S(500\mu\textrm{m})\ge100$ mJy. The HerBS survey is based on a $S(500\mu\textrm{m})>80$ mJy flux cut and a photometric redshift $z_\textrm{phot}>2$ selection. 
All sources with spectroscopic redshifts already available and known blazars were excluded from the list of potential targets, thus resulting in the above mentioned 137 objects observed with NOEMA. We defer to Paper I for further details about the $z$-GAL source selection and observations, as well as for a description of the data calibration and reduction. The measurements of lines and continuum fluxes are described in Paper I and Paper II, respectively.

The NOEMA data revealed that several of these {\it Herschel}-selected targets consist of multiple components, that are detected in the dust continuum and emission lines. Taking into account all the multiple sources identified in the targeted fields, spectroscopic redshifts were measured for 165 individual objects in the range $0.8<z<6.5$ (Paper I).

\subsection{Spectral lines}\label{sect:data_lines}

The spectral emission lines detected in the $z$-GAL NOEMA spectra were fitted with a Gaussian profile with a simple least squares method. Up to four spectral lines were fitted simultaneously for a given source, thus leading to very precise redshift measurements (with typical errors of few $10^{-4}$ in redshift). For each source, all detected lines were assumed to have the same velocity width.
When single Gaussian profiles were not sufficient to reproduce the observed line profile, double Gaussians were adopted. In this case, the algorithm also assumed that the velocity spacing between the fitted Gaussian profiles was the same for all detected lines. 
The determination of the redshifts, line intensities, line widths, velocity separations, and their respective errors, was based solely on the rest frequencies of the lines and on the S/N with which they were detected. 

Paper I presents the line properties measured for all 137 $z$-GAL targets, including their sky coordinates, spectroscopic redshift, widths and integrated fluxes. For the majority of sources (85\%) at least two $^{12}$CO lines are detected; 21\% of the sources benefit from the detection of three spectral lines, including $\rm ^{12}CO$ and other species, such as \hto\ and [C{\small I}]; finally 8\% of the sources benefit from the detection of three $\rm ^{12}CO$ transitions (Sect. 3.1 of Paper I). 
The spectra and maps of each $z$-GAL target are presented in the Appendix of Paper I.

\subsection{Continuum}\label{sect:data_continuum}

Paper II presents the NOEMA continuum catalogue of the $z$-GAL survey. Continuum fluxes were measured through polygonal apertures from cleaned continuum maps. Flux statistical uncertainties were computed rescaling the map noise to the effective extraction aperture size. 

The spectral energy distributions (SEDs) of the $z$-GAL sources were modelled in Paper II. A modified black body (MBB) in its general form and in the optically thin approximation was adopted. The products of this analysis are the dust mass $M_\textrm{dust}$, temperature $T_\textrm{dust}$ and emissivity index $\beta_\textrm{dust}$ of the sources, as well as their IR luminosity integrated between 50 and 1000 $\mu$m. 
To ensure an easy comparison to the previous works found in the literature, Paper II adopts the optically thin solution as reference and discuss the consequences of this approximation by comparing the results to those obtained with MBB in its general form. Therefore the same choice is made here. 

The IR luminosities computed by integrating the MBB model are here used to estimate the SFR of the sources and to normalise the line fluxes when computing their median SLED. To this aim, the 50-1000 $\mu$m luminosities derived in Paper II need to be transformed into the total IR luminosity $L_\textrm{IR}(8-1000\ \mu\textrm{m})$. 
By integrating the SED templates by \citet[][star-forming galaxies only]{berta2013}, we derive a median ratio of $L(50-1000\ \mu\textrm{m})$ to $L_\textrm{IR}$ of 0.7 with a median absolute deviation of 0.1. We adopt this value throughout this study.
The choice of this library is driven mainly by the fact that it was built on a multicolour study of {\it Herschel} galaxies. It is certainly not exclusive and other template libraries could be used. We note however that we specifically avoid luminosity-dependent template libraries because a significant fraction of our sources is likely lensed (Paper IV) and their intrinsic luminosity is not known yet.


\section{Properties of molecular gas}\label{sect:properties_mol_gas}

The rich $z$-GAL lines catalogue (Paper I) covers the $^{12}$CO transitions from $J_\textrm{up}=2$ to $J_\textrm{up}=8$, with at least two $^{12}$CO transitions available for most sources and, for a few, also the \hto\ or [C{\small I}] lines (Sects. \ref{sect:data_lines}, \ref{sect:H2O_line}, and \ref{sect:ci}). In this Section we present the main properties of the detected $\rm ^{12}CO$ and water emission lines. We compare them to those of different samples found in the literature, with the goal to understand the process that dominates the line emission of our sources and their nature. The $\rm ^{12}CO$ line ratios of individual sources are studied in Sect. \ref{sect:CO_lines}; the average $\rm ^{12}CO$ SLED is presented in Sect. \ref{sect:avg_sled}; the SLEDs of sources with at least three $\rm ^{12}CO$ transitions available are modelled in Sect. \ref{sect:sleds_lvg}; and finally the properties of water lines are discussed in Sect. \ref{sect:H2O_line}.

The observed intensity of a spectral emission line can be translated into its luminosity (in units of $\rm K \, km \, s^{-1} pc^2$) using the standard relation \citep[e.g.][]{Solomon-VandenBout2005}:
\begin{equation}\label{eq:Lprime_line}
L^\prime_\textrm{line}= 3.25\times10^7 \,  S_\textrm{line} \Delta V \times \frac{D_L^2}{(1+z)} \frac{1}{\nu_\textrm{rest}^2} \textrm{,}
\end{equation}
where $S_\textrm{line} \Delta V$ is the velocity-integrated line intensity in units of $\rm Jy \, km \, s^{-1}$, $\nu_\textrm{rest}$ is the line rest frequency in GHz, and $D_L$ is the luminosity distance in Mpc.

\subsection{CO lines ratios}\label{sect:CO_lines}

\begin{table*}[!t]
\caption{Median $L^\prime(^{12}\textrm{CO})$ line luminosity ratios and $I(^{12}\textrm{CO})$ line intensity ratios, measured for the full $z$-GAL + pilot sample. The number $N$ of sources participating in the median for each transition is listed in column 2. When only one source is available, we use the actual ratio and the uncertainty computed with standard error propagation, when only two sources are available we use the weighted average. }
\label{tab:line_ratios}
\centering
\begin{tabular}{lcccccc|lccc}
\hline
\hline
$L^\prime$ ratio & $N$ & median & m.a.d. & CW13 & B20 & H21 & $I$ ratio & median & m.a.d. \\ 
\hline
$r_{32/21}$ & 15 & 0.93 & 0.15 & 0.78 & 0.90$\pm$0.23 & 0.78$\pm$0.15 & $R_{32/21}$ & 2.10 & 0.34   \\
$r_{43/21}$ & 22 & 0.56 & 0.18 & 0.54 & 0.75$\pm$0.21 & 0.59$\pm$0.17 & $R_{43/21}$ & 2.23 & 0.74   \\
$r_{43/32}$ & 52 & 0.81 & 0.13 & 0.70 & 0.83$\pm$0.23 & 0.75$\pm$0.24 & $R_{43/32}$ & 1.45 & 0.23   \\
$r_{54/32}$ & 47 & 0.60 & 0.23 & 0.59 & 0.64$\pm$0.19 & 0.54$\pm$0.24 & $R_{54/32}$ & 1.68 & 0.63   \\
$r_{65/32}$ &  4 & 0.54 & 0.10 & --   & 0.42$\pm$0.15 & 0.36$\pm$0.21 & $R_{65/32}$ & 2.17 & 0.29   \\
$r_{54/43}$ & 13 & 0.64 & 0.21 & --   & 0.78$\pm$0.25 & 0.71$\pm$0.35 & $R_{54/43}$ & 1.00 & 0.32   \\
$r_{65/43}$ & 15 & 0.68 & 0.15 & --   & 0.51$\pm$0.19 & 0.48$\pm$0.30 & $R_{65/43}$ & 1.54 & 0.35   \\
$r_{76/43}$ &  5 & 0.44 & 0.07 & --   & 0.25$\pm$0.13 & 0.33$\pm$0.25 & $R_{76/43}$ & 1.35 & 0.23   \\
$r_{65/54}$ &  2 & 2.08 & 0.98 & --   & 0.66$\pm$0.26 & 0.68$\pm$0.47 & $R_{65/54}$ & 2.99 & 1.99   \\
$r_{76/54}$ &  3 & 1.29 & 0.44 & --   & 0.32$\pm$0.17 & 0.46$\pm$0.37 & $R_{76/54}$ & 2.52 & 0.86   \\
$r_{87/54}$ &  1 & 0.24 & 0.05 & --   & 0.12$\pm$0.11 & 0.30$\pm$0.27 & $R_{87/54}$ & 0.62 & 0.07   \\
\hline
\end{tabular}
\tablefoot{The median absolute deviation of a sample of values $x$ is defined as $\textrm{m.a.d.}\left(x\right)=\textrm{median}\left(\left|x-\textrm{median}\left(x\right)\right|\right)$.
The 5th, 6th and 7th columns report the same $L^\prime$ ratios derived from the Tables in \citet[][CW13]{Carilli-Walter2013}, \citet[][B20]{boogaard2020}, and \citet[][H21]{harrington2021}.}
\end{table*}

\begin{figure*}[!t]
\centering
\rotatebox{-90}{\includegraphics[height=0.47\textwidth]{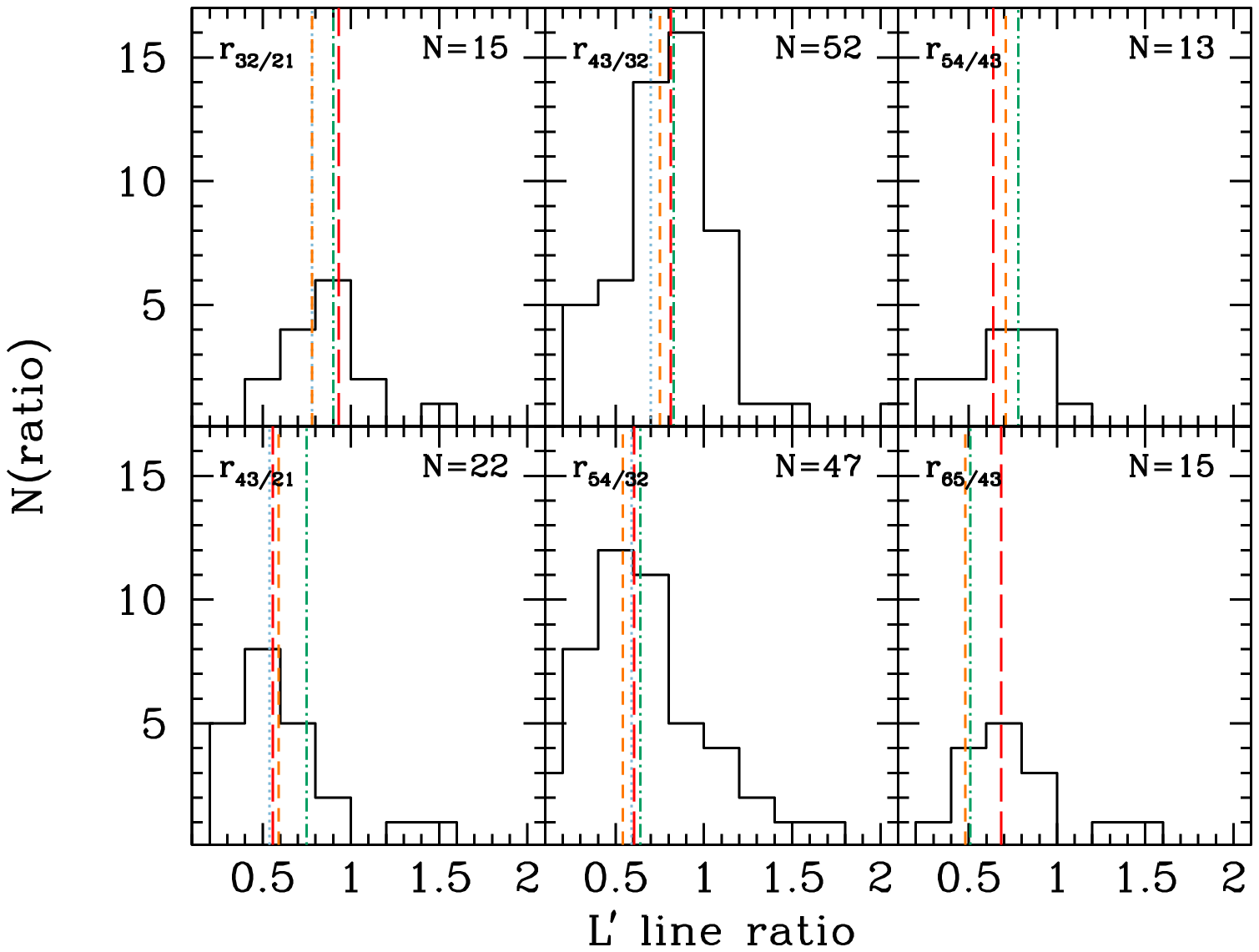}}
\rotatebox{-90}{\includegraphics[height=0.47\textwidth]{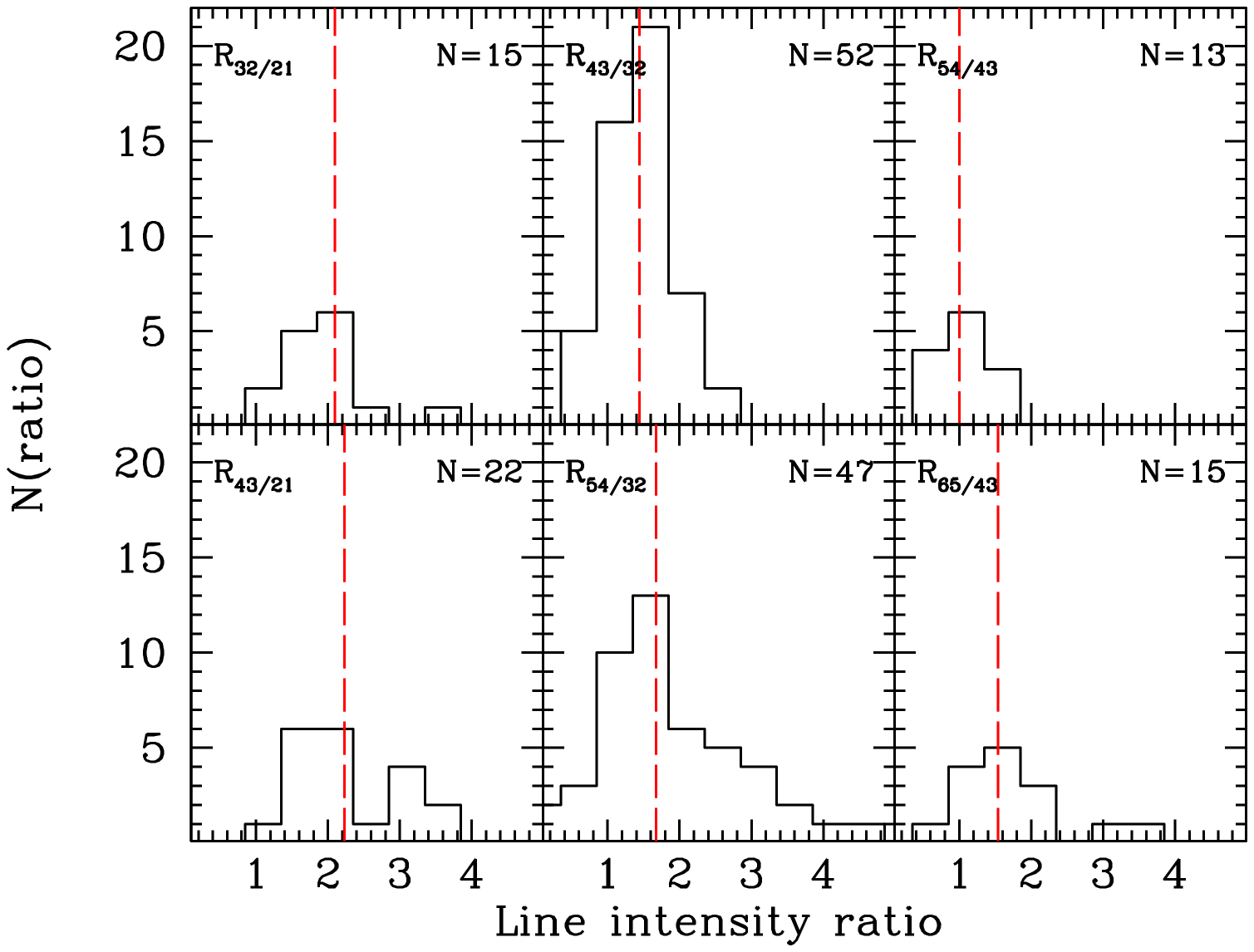}}
\caption{$\rm ^{12}CO$ line ratios. {\em Left panels:} $L^\prime$ line luminosity ratios, $r_{ij/lm}$ between the transitions $^{12}\textrm{CO}(i-j)/^{12}\textrm{CO}(l-m)$. The red long-dashed vertical lines mark the median $z$-GAL ratios (Tab. \ref{tab:line_ratios}). The dotted light-blue vertical lines represent the ratios reported for SMGs by CW13; the short dash orange lines the H21 SMG ratios; and the dot-dash green lines the B20 line ratios of less luminous galaxies, complementary to the $z$-GAL sample. {\em Right panels}: Observed intensity $I$ ratios, $R_{ij/lm}$; the red long-dashed vertical lines mark the median values.}
\label{fig:Lprime_CO_line_ratios}
\end{figure*}

Table \ref{tab:LprimeCO_all} lists the values of $L^\prime_\textrm{CO}$ obtained for all the detected $\rm ^{12}CO$ transitions of the $z$-GAL sources, as derived from the catalogue presented in Paper I. 
Table \ref{tab:line_ratios} lists the median $L^\prime_\textrm{CO}$ line luminosity ratios, $r_{ij/lm}$, and the median $I_\textrm{CO}$ line intensity ratios, $R_{ij/lm}$, representing the ratios between the transitions $^{12}\textrm{CO}(i-j)/^{12}\textrm{CO}(l-m)$. For comparison, we derive the corresponding $L^\prime_\textrm{CO}$ ratios from the Tables by \citet[][CW13]{Carilli-Walter2013}, \citet[][B20, eight galaxies in the redshift range $z=2.0-2.7$]{boogaard2020}, and \citet[][H21]{harrington2021}. We apply standard error propagation to derive the uncertainties associated with the B20 and H21 ratios. Except the cases with less than ten sources available, the $z$-GAL ratios are comparable to those of sub-millimetre galaxies (SMGs) by CW13 and H21. We attribute the discrepancy of the remaining line ratios to small number statistics. The B20 sample finally consists of less luminous galaxies, with a median IR luminosity of $L_\textrm{IR}\sim 8\times 10^{11}$ L$_\odot$, and is therefore complementary to $z$-GAL.

Figure \ref{fig:Lprime_CO_line_ratios} shows the distribution of the computed $L^\prime$ and $I$ ratios for those transitions with at least ten sources available. 
Possible trends of the $\rm ^{12} CO$ line ratios as a function of redshift or FWHM are investigated but none are found.

For completeness sake, we recall that \citet{stanley2022} observed the 11 $z$-GAL pilot targets \citep{neri2020} with the VLA, detecting   the $\rm ^{12} CO(1-0)$ transition in all of them. These authors combine their $\rm ^{12} CO(1-0)$ measurements to the $(3-2)$, $(4-3)$, and $(5-4)$ by \citet{neri2020} and report median ratios: $\textrm{r}_{32/10}=0.69$, $\textrm{r}_{43/10}=0.64$ and $\textrm{r}_{54/10}=0.74$. 
These ratios are to be compared to those measured by H21 for {\it Planck} lensed galaxies, $\textrm{r}_{32/10}=0.69\pm0.12$, $\textrm{r}_{43/10}=0.52\pm0.14$, $\textrm{r}_{54/10}=0.37\pm0.15$, and those collected by CW13 for SMGs, $\textrm{r}_{32/10}=0.66$, $\textrm{r}_{43/10}=0.46$, and $\textrm{r}_{54/10}=0.39$. The results by \citet{stanley2022} indicate a higher $\rm ^{12}CO$ excitation of the $z$-GAL pilot sources with respect to CW13 and H21.


\begin{figure}[!t]
\centering
\rotatebox{-90}{\includegraphics[height=0.47\textwidth]{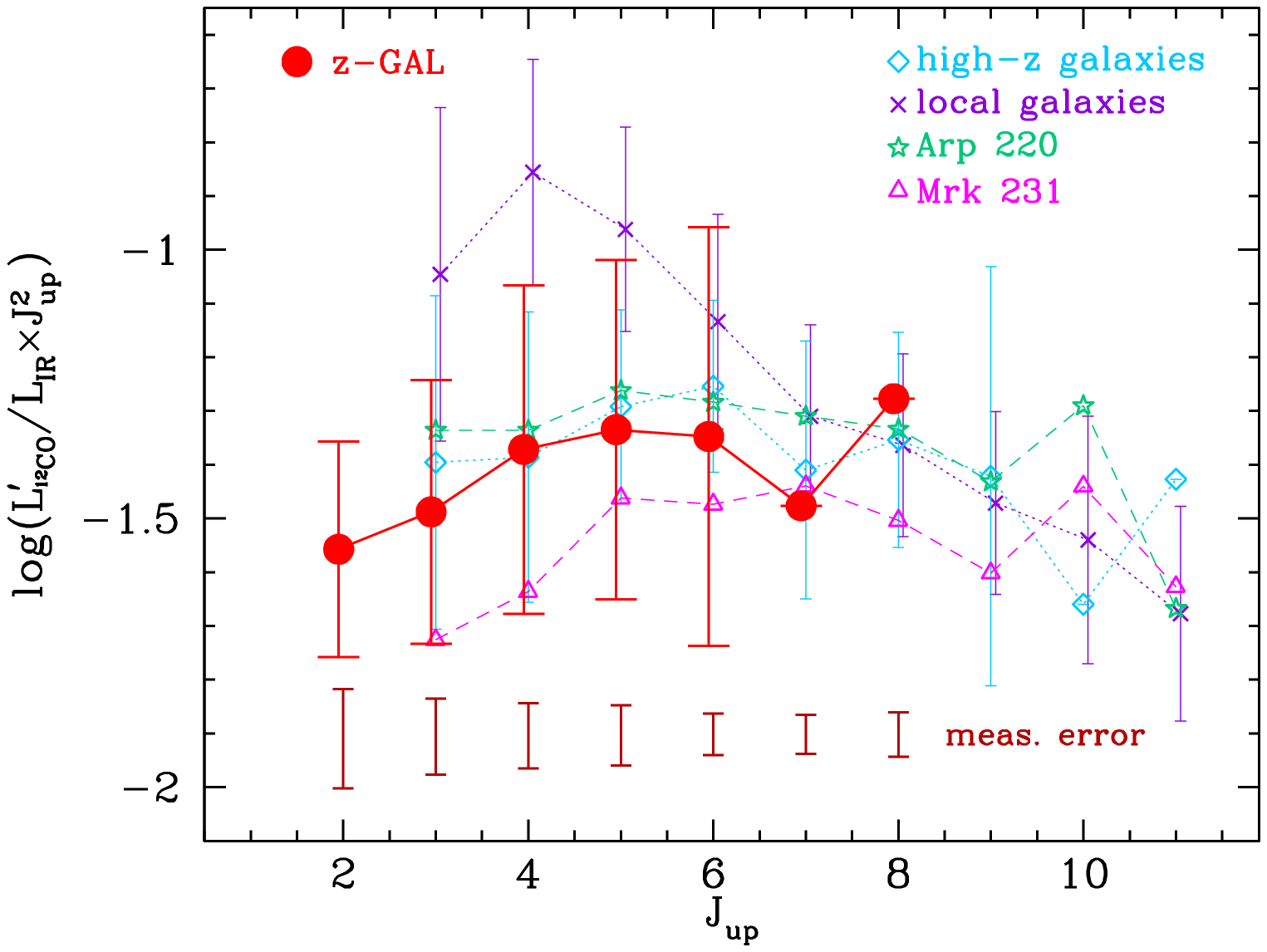}}
\caption{$L_\textrm{IR}$-normalised average SLED of the $z$-GAL sources, compared to high-$z$ SMGs \citep{Carilli-Walter2013,yang2017}, local star-forming galaxies \citep{liu2015}, Arp220 \citep{rangwala2011} and Mrk231 \citep{vanderwerf2010}. In the case of $z$-GAL data points, if the volume of the $L^\prime_\textrm{CO}/L_\textrm{IR}\times J_\textrm{up}^2$ values is insufficient for a normal distribution fitting, no uncertainty is shown. Error bars are computed as the Gaussian dispersion of the $L^\prime_\textrm{CO}/L_\textrm{IR}\times J_\textrm{up}^2$ distribution of each transition (not available for $J_\textrm{up}=7$ and 8 because of the small statistics). For comparison, the dark-red error bars on the bottom represent the median uncertainty on $L^\prime_\textrm{CO}/L_\textrm{IR}\times J_\textrm{up}^2$ based on measurements errors only, as obtained via standard error propagation.}
\label{fig:LpCO_LIR_SLED}
\end{figure}

\subsection{Average $\rm ^{12}CO$ SLED}\label{sect:avg_sled}

The average $\rm ^{12}CO$ SLED of the $z$-GAL sample has been computed as the Gaussian average and deviation of the ratio
$L^\prime_\textrm{CO}/L_\textrm{IR}$ of each detected transition multiplied by $J_\textrm{up}^2$  \citep[see][]{yang2017}.
In Fig. \ref{fig:LpCO_LIR_SLED}, the result is shown as a function of $J_\textrm{up}$ and compared to other $L_\textrm{IR}$-normalised $\rm ^{12}CO$ SLEDs found in the literature.   

The $z$-GAL average SLED is consistent with high-redshift SMGs \citep{Carilli-Walter2013,yang2017} within the error bars and similar to the local star-formation-dominated ULIRG Arp220 \citep{rangwala2011}. The flatness of these SLEDs demonstrates that in such objects the low-excitation CO component (peaking at $J_\textrm{up}=3-4$) is marginal, at odds with local star-forming galaxies \citep{liu2015}. The SLED of the AGN-powered ULIRG Mrk 231 \citep{vanderwerf2010} is also flat, but its CO/IR luminosity ratio is overall significantly lower than in SF-dominated galaxies.


\subsection{LVG analysis of individual $\rm ^{12}CO$ SLEDs}\label{sect:sleds_lvg}

\begin{table*}[!ht]
\caption{Results of LVG modelling of the $^{12}$CO SLEDs of $z$-GAL individual sources with at least three $^{12}$CO transition available. The quoted uncertainties are 1$\sigma$.}
\label{tab:sleds_lvg_results}
\centering
\begin{tabular}{lcccc}
\hline
\hline
Source  & $\log(n_{\textrm{H}_2})$ &      $\log(T_\textrm{kin})$  &  $\log(N_\textrm{CO}/dv)$ &           $\log(P_\textrm{th})$ \T \\
	&        $\log(\textrm{cm}^{-3})$ & $\log(\textrm{K})$ & $\log(\textrm{cm}^{-2}\textrm{ km}^{-1}\textrm{ s})$ & $\log(\textrm{K cm}^{-3})$ \B \\
\hline
HeLMS-36      &  3.6$^{+0.8}_{-0.9}$   &   2.4$^{+0.4}_{-0.5}$  &     17.4$^{+0.8}_{-0.9}$    &    6.0$^{+0.6}_{-0.9}$   \T \\
HeLMS-38      &  3.3$^{+0.8}_{-0.7}$   &   2.0$^{+0.6}_{-0.5}$  &     16.3$^{+0.8}_{-0.5}$    &    5.5$^{+0.3}_{-0.6}$   \T \\
HeLMS-48      &  4.0$^{+0.9}_{-0.8}$   &   2.0$^{+0.6}_{-0.4}$  &     16.2$^{+0.9}_{-0.5}$    &    6.2$^{+0.4}_{-0.5}$   \T \\
HerS-14       &  3.8$^{+0.5}_{-0.6}$   &   2.4$^{+0.4}_{-0.4}$  &     16.8$^{+0.9}_{-0.8}$    &    6.3$^{+0.3}_{-0.7}$   \T \\ 
HerBS-61      &  3.3$^{+0.8}_{-0.8}$   &   2.2$^{+0.5}_{-0.5}$  &     17.4$^{+0.7}_{-0.8}$    &    5.7$^{+0.6}_{-0.8}$   \T \\
HerBS-78      &  3.9$^{+0.8}_{-0.7}$   &   2.2$^{+0.5}_{-0.5}$  &     17.5$^{+0.7}_{-0.8}$    &    5.7$^{+0.6}_{-0.8}$   \T \\
HerBS-193     &  3.7$^{+0.7}_{-0.9}$   &   2.3$^{+0.5}_{-0.5}$  &     16.5$^{+1.0}_{-0.7}$    &    6.2$^{+0.3}_{-1.0}$   \T \B \\
\hline
\end{tabular}
\end{table*}

\begin{figure*}[!t]
\centering
\includegraphics[width=0.4\textwidth]{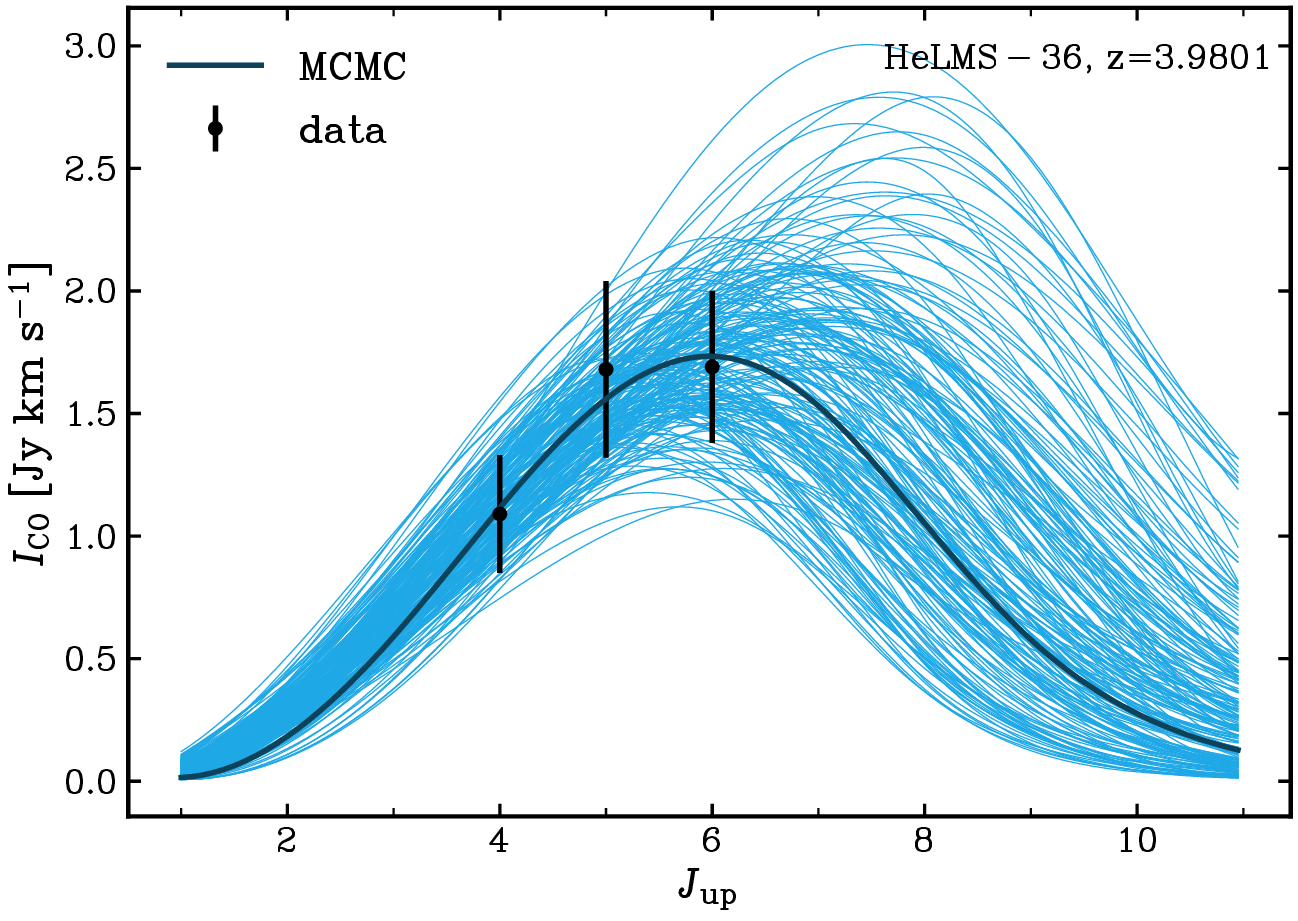}	
\hspace{0.05\textwidth}
\includegraphics[width=0.4\textwidth]{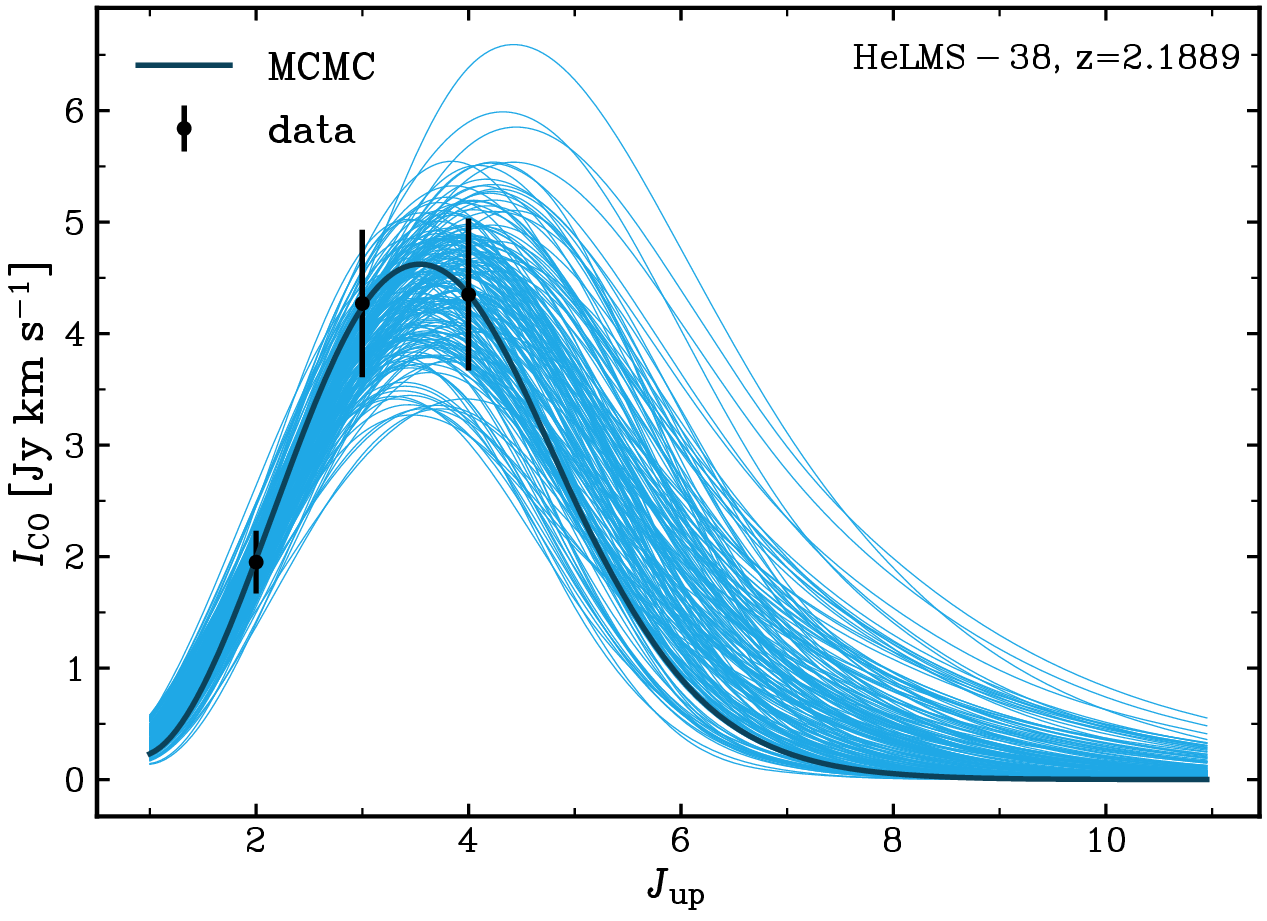}\\
\includegraphics[width=0.4\textwidth]{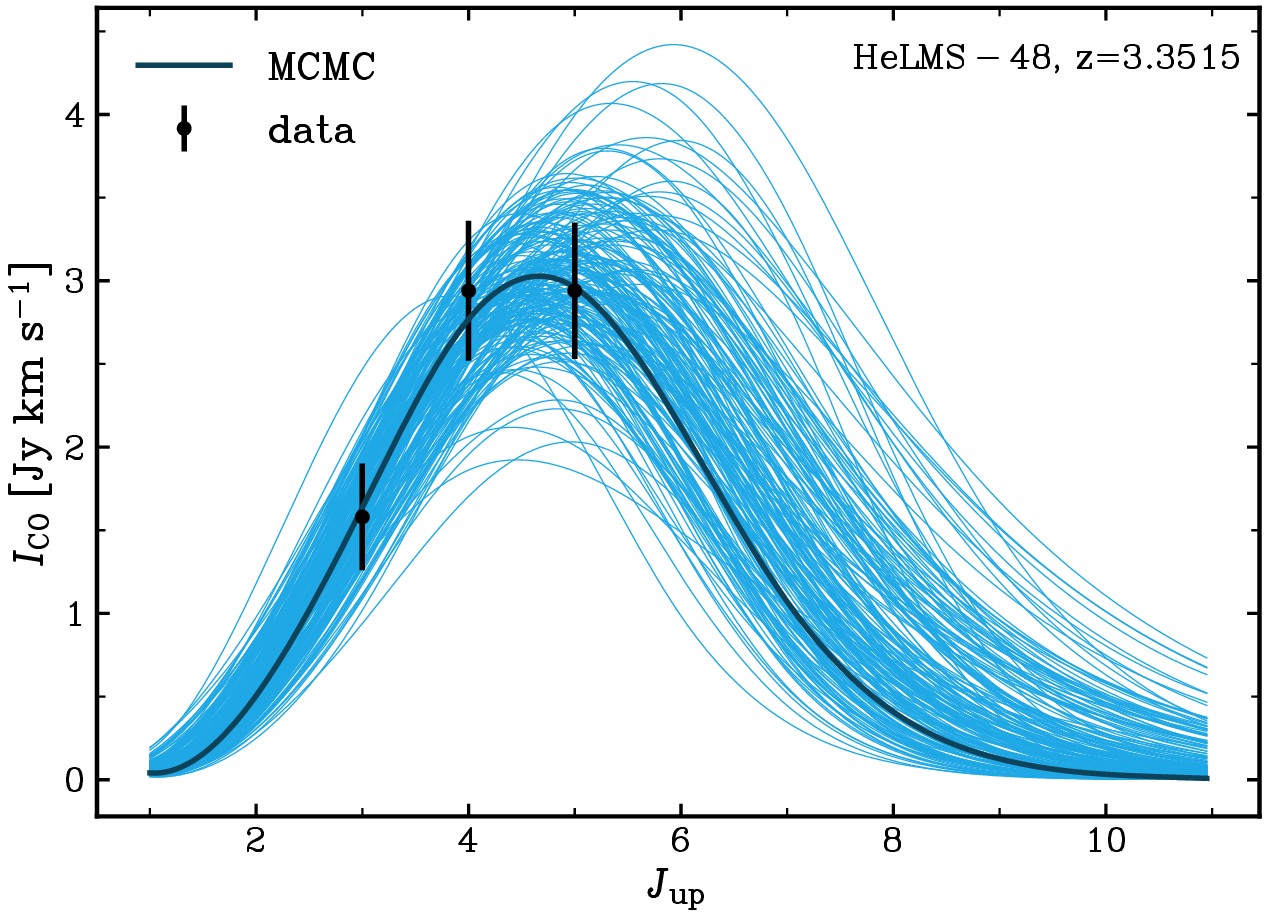}
\hspace{0.05\textwidth}
\includegraphics[width=0.4\textwidth]{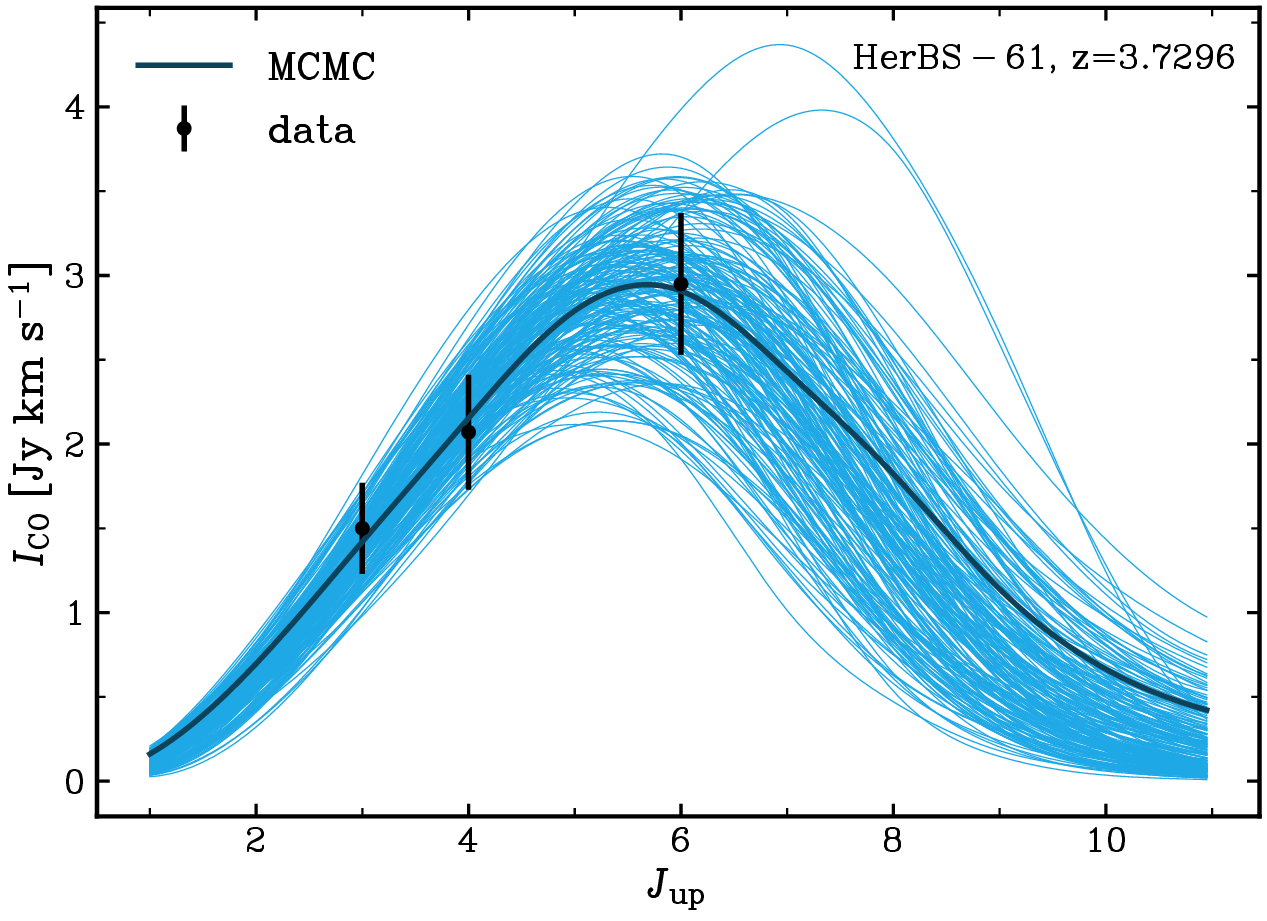}\\
\includegraphics[width=0.4\textwidth]{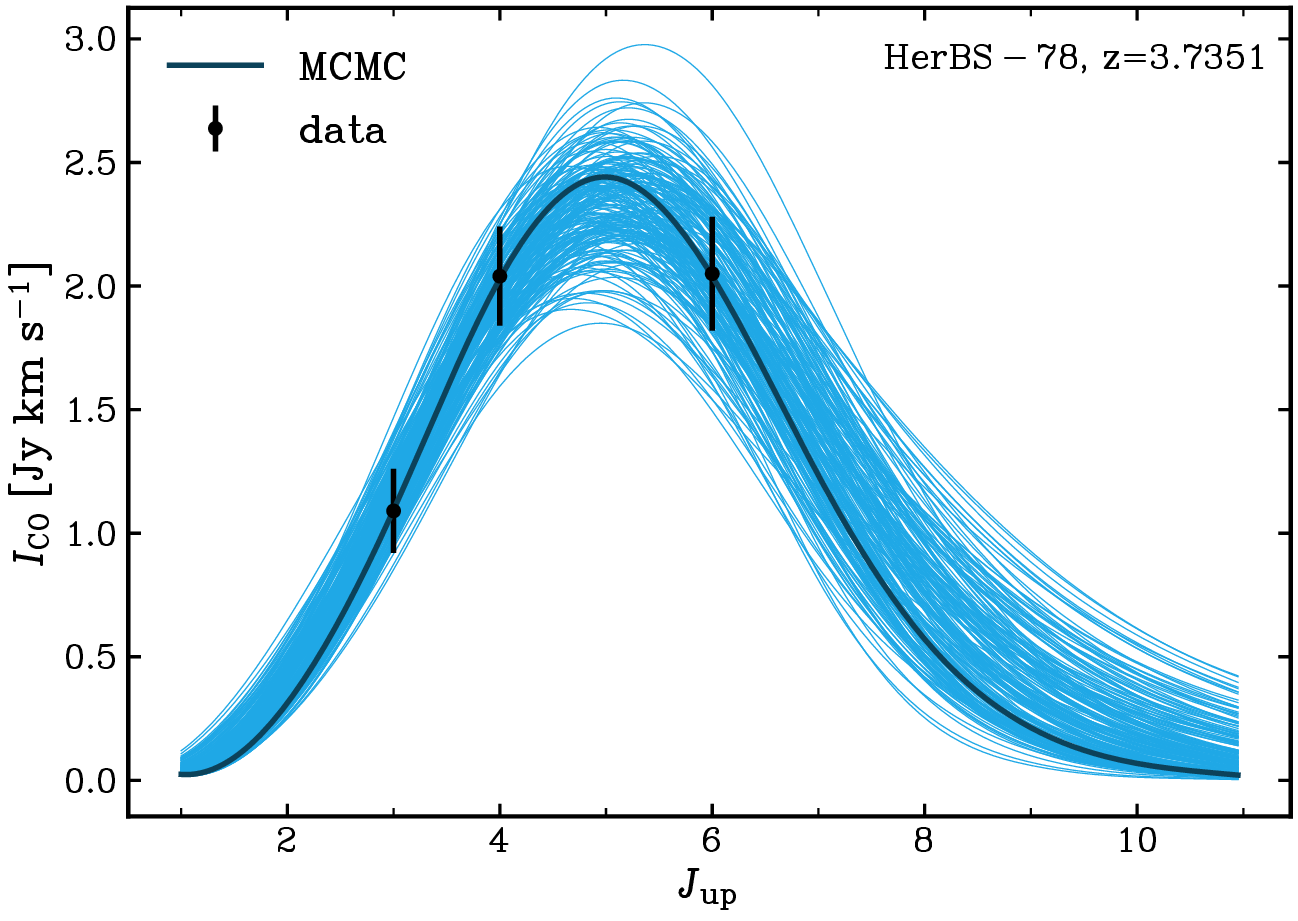}
\hspace{0.05\textwidth}
\includegraphics[width=0.4\textwidth]{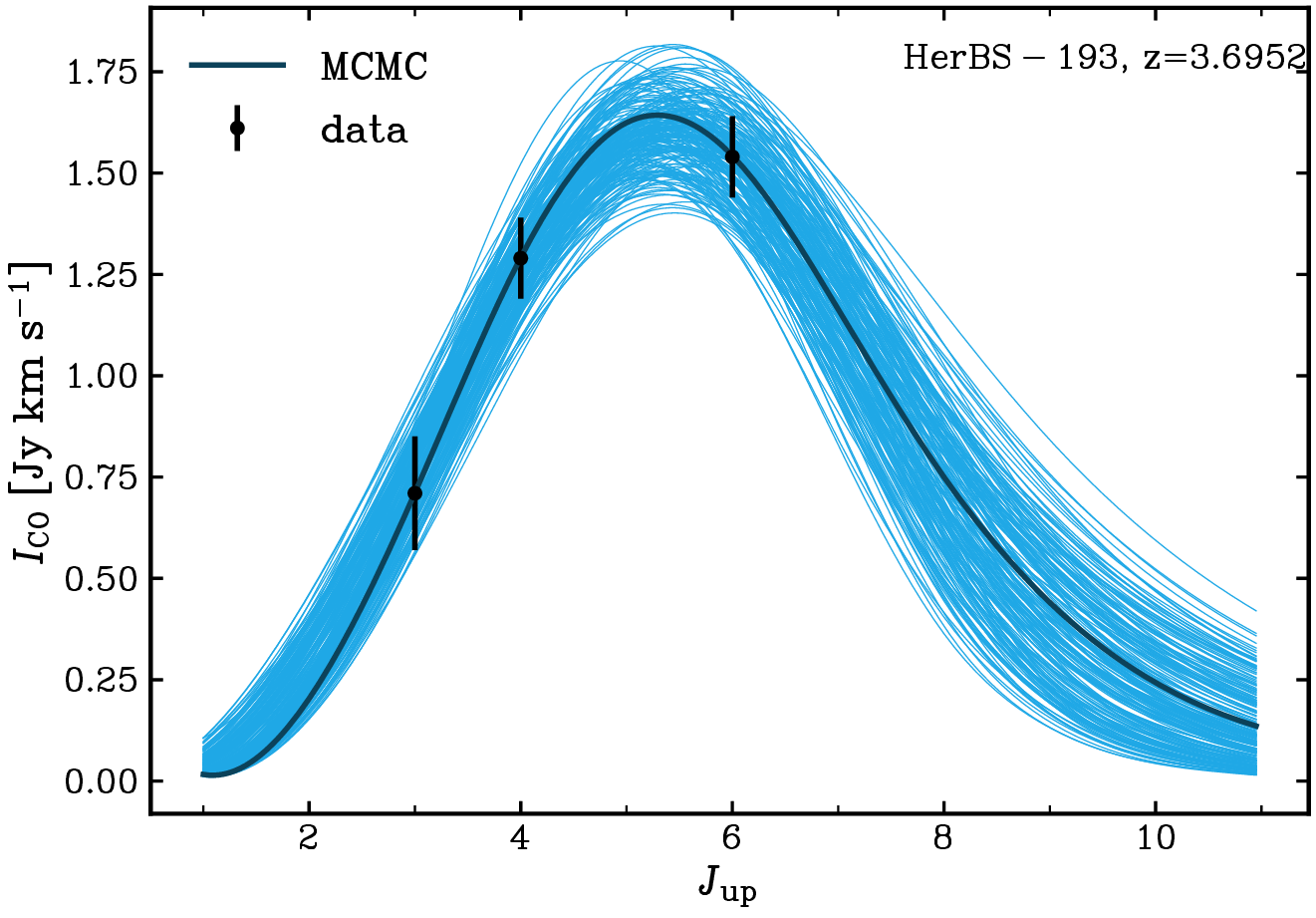}\\
\includegraphics[width=0.4\textwidth]{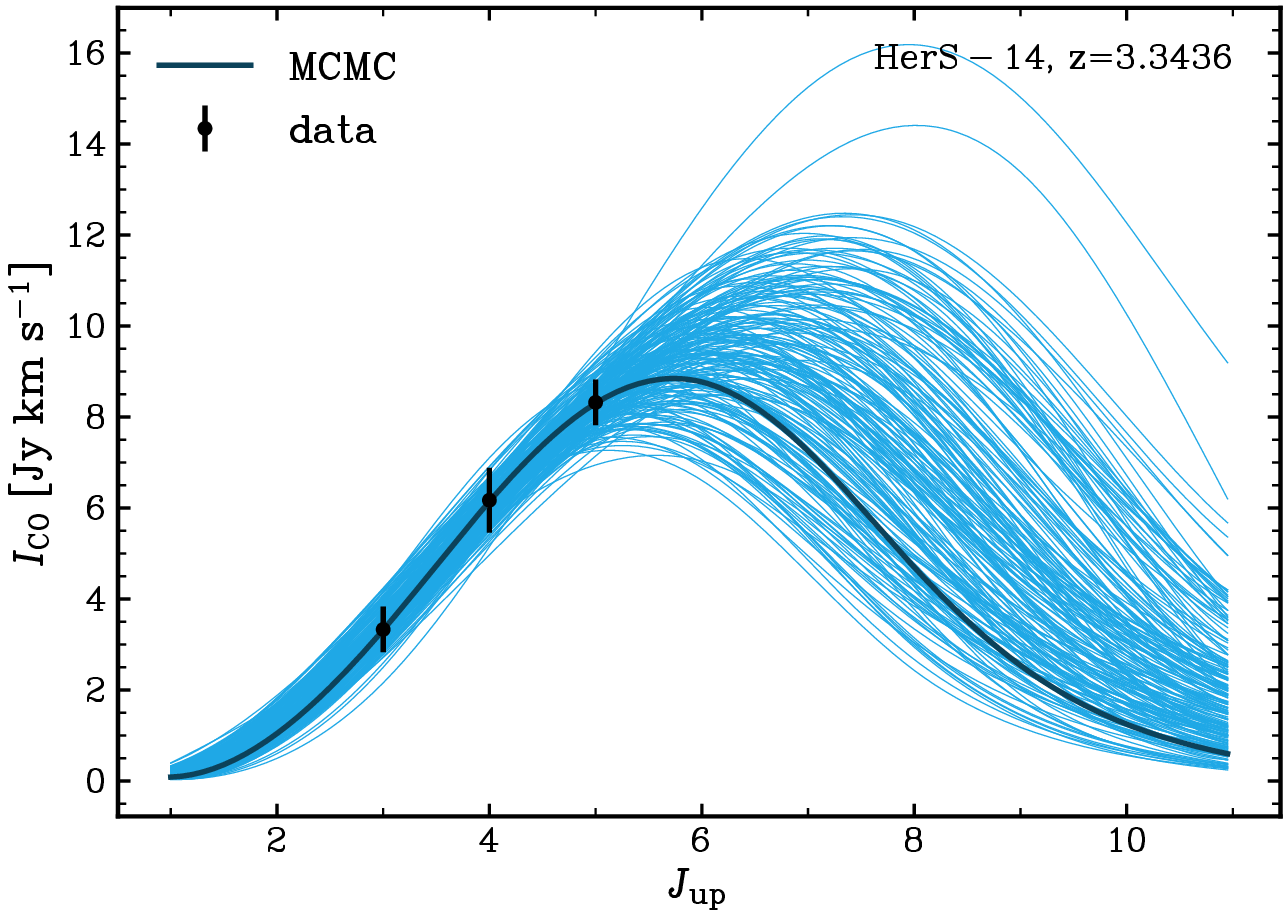}
\caption{Results of LVG fit of the $\rm ^{12}CO$ SLEDs. The best fit model is plotted in black. The blue thin lines represent all models within the $\pm1\sigma$ range of the posterior distribution around the best fit.}
\label{fig:SLED_LVG}
\end{figure*}

We modelled the $\rm ^{12}CO$ excitation and the physical conditions of the molecular gas using the large velocity gradient (LVG) statistical equilibrium method \citep[e.g.][]{sobolev1960} for seven $z$-GAL sources with at least three $\rm ^{12}CO$ transitions detected. Two further sources (HerBS-43b and HerBS-58) are described by \citet{stanley2022}. We adopted a one-dimensional (1D) non-LTE radiative transfer code \texttt{RADEX} \citep{vadertak2007}, with an escape probability of $\beta=(1-e^{-\tau})/\tau$ derived from an expanding sphere geometry. The $\rm ^{12}CO$ collisional data are from the LAMDA database \citep{schoier2005}. With the same MCMC (Monte Carlo Markov Chain) approach used by \citet{yang2017}, we derive the posterior distributions of the kinetic temperature of the molecular gas ($T_{\mathrm{kin}}$), the volume density ($n_{\mathrm{H_2}}$), the column density of $\rm ^{12}CO$ per unit velocity gradient ($N_{\mathrm{CO}}/dV$), and the solid angle ($\Omega_{\mathrm{app}}$) of the source. Assuming a similar filling factor and magnification across the $J$ transitions of the $\rm ^{12}CO$,  the overall shape of the $\rm ^{12}CO$ SLEDs only depends on $T_{\mathrm{kin}}$, $n_{\mathrm{H_2}}$, and $N_{\mathrm{CO}}/dV$, and scales with $\Omega_{\mathrm{app}}$ (also including the magnification factors). Therefore we focus only on the parameters $T_{\mathrm{kin}}$, $n_{\mathrm{H_2}}$ and $N_{\mathrm{CO}}/dV$.

As the transitions of the $\rm ^{12}CO$ lines observed are $J_\textrm{up}\le7$, there are insufficient data to constrain the highly excited molecular gas component, which usually peaks around $J_\textrm{up}=8$ \citep{yang2017}. Therefore we assume that our galaxies are similar to other high-redshift SMGs, in which the $\rm ^{12}CO$ excitation is dominated by two components peaking around $J=6$ and 8, respectively \citep{yang2017,canameras2018}. Accordingly, to better constrain the posteriors, we used slightly tighter boundaries for the flat priors of $n_{\mathrm{H_2}}$ and $N_{\mathrm{CO}}/dV$ compared to \citet{yang2017}, while other priors are unchanged. Taking the values of the parameters from statistically studied SMG samples \citep{yang2017,canameras2018}, we have chosen flat priors of $\log(n_{\mathrm{H_2}}/\textrm{cm}^{-2}$) between 2.0 and 5.5 and $\log(N_{\mathrm{CO}}/dV/\textrm{cm}^{-2} (\textrm{km/s})^{-1}$) between 15.5 and 18.5. Similarly, we also limited the range of the thermal pressure $P_\mathrm{th}$ (defined by $P_\mathrm{th}\equiv n_{\mathrm{H_2}} \times T_{\mathrm{kin}}$) to be within $10^4$ and $10^7$ K cm$^{-3}$. 

A total of 100\,000 points of the solutions have been explored in the parameters space, with two hundred walkers and five hundred iterations after the one hundred burn-in runs. 
Figure \ref{fig:SLED_LVG} shows the observed SLED and the models. The results are reported in Tab. \ref{tab:sleds_lvg_results}, indicating the $\pm$ 1$\sigma$ values and the median of the posteriors. 
Most of the values of the molecular gas temperature are in the range from 100 to 250 K, while the density varies from $10^{3.3}$ to $10^{4.0}$ cm$^{-3}$. These values are consistent with the ones found in high-redshift SMGs \citep[][$\log(n_{\textrm{H}_2}/\textrm{cm}^{-3})\simeq2.5-4.1$, $T_\textrm{kin}\simeq20-750$ K]{yang2017,canameras2018,harrington2021} and with those of the Pilot Programme \citep[][$\log(n_{\textrm{H}_2}/\textrm{cm}^{-3})=2.5-3.9$, $T_\textrm{kin}=100-200$ K]{stanley2022}.


\subsection{Water lines}\label{sect:H2O_line}

\begin{figure}[!t]
\centering
\includegraphics[width=0.45\textwidth]{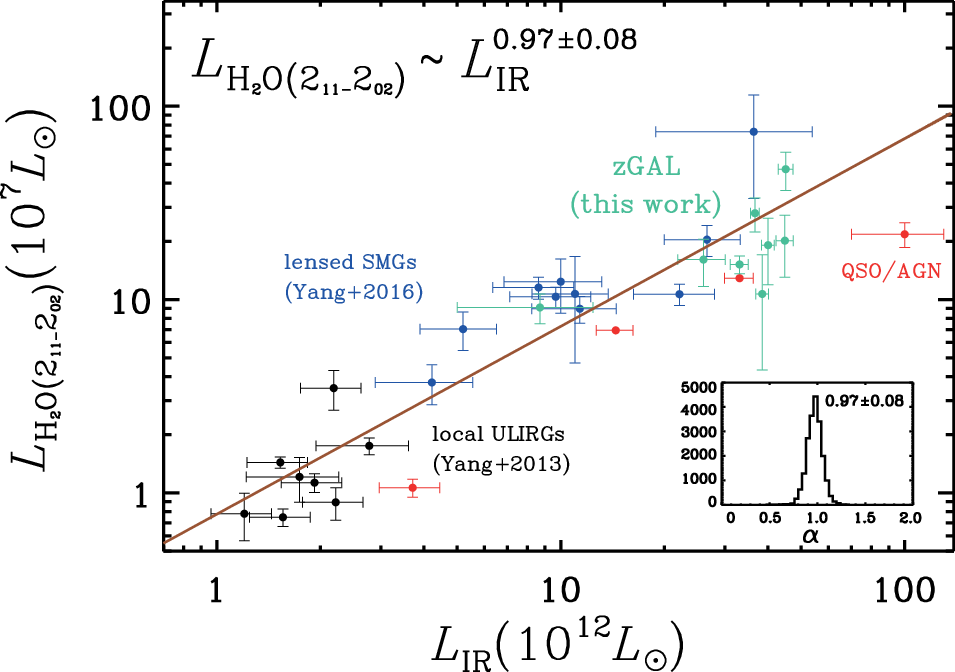}
\caption{Correlation between $L_{\rm IR}$ and $L_{\rm H_2O(2_{11}-2_{02})}$ 
for local (black symbols), high-redshift star-forming 
SMGs (blue symbols), and QSO/AGN (red). $z$-GAL sources are marked 
in green. The fitting of the correlation is represented by the solid orange line. 
The posterior distribution of the slope is shown in the inset.
The plot is adapted from \cite[][see references therein]{Yang2016}. The new fit is performed by including the $z$-GAL sources (green symbols) as well as the sources from the literature.}
\label{fig:water_LIR}
\end{figure}

Water is one of the most abundant molecules after $\rm H_2$ and 
CO in the gaseous ISM \cite[e.g.][]{vd2013}. 
The emission and absorption of the \hto\ lines trace a variety of physical 
processes such as shocks, collisions and radiative pumping. Therefore 
it probes the physical conditions of the 
inter-stellar medium (ISM) in both local \cite[e.g.][]{Gonzalez-Alfonso2014} and high-redshift 
galaxies \cite[e.g.][]{Omont2013, Yang2016, Yang2020, jarugula2019}. 
Being $E_\mathrm{upper}\gtrapprox130$~K, \hto\ is primarily 
excited through absorption of far-IR photons emitted by warm dust in dense regions. Only a small contribution from collisional excitation is expected, therefore it is a useful diagnostic of the far-IR radiation field \cite[e.g.][]{ga22}. Observationally, 
these sub-millimetre \hto\ lines are the second strongest molecular emission 
lines after CO \cite[e.g.][]{Yang2013}. 

In the $z$-GAL sample, we have detected seven sources in the para-$\rm H_2O(2_{11}-2_{02})$ line ($\nu_\textrm{rest}=752$ GHz) with $E_\textrm{upper}=137$~K: HeLMS-17 W, HerBS-38 NE, 83, 177, 179, 185 (see Tables in the Appendix of Paper I), and finally HerBS-154 from the Pilot Programme \citep{neri2020}. 
We note that the $z=6.5678$ HerBS-38 NE source lies in a field with two other sources at $z=2.4775$ and 2.4158, labelled HerBS-38 SE and W (Paper I). Because of blending in the {\it Herschel} bands, it is not possible to estimate its IR luminosity, therefore this source is not included in this piece of analysis.

The line widths of these water lines are similar to those of the CO lines, similarly to the findings of other studies \cite[e.g.][]{Omont2013, Yang2016}, suggesting that they are emitted by the same star-forming regions across the sources. 
The observed line luminosities of the para-$\rm H_2O(2_{11}-2_{02})$ lines are in the range $\mu L_{\rm H_2O}=1.1$ to $4.7\times 10^{8}$ L$_\odot$,
placing them amongst the brightest \hto\ emitters identified to date (modulo gravitational lensing magnification). 

In Fig. \ref{fig:water_LIR} the correlation between the $\rm H_2O(2_{11}-2_{02})$ luminosity and $L_\textrm{IR}$ of the $z$-GAL sources is presented together with results from previous studies.
Including the $z$-GAL sources, the slope of this correlation is found to be $0.97\pm0.08$, slightly shallower than the relation $L_{\rm H_2O(2_{11}-2_{02})} \sim L_{\rm IR}^{1.16\pm0.13}$ reported by \citet{Yang2016}, but still in good agreement within the uncertainties. 
The average $L_{\rm H_2O(2_{11}-2_{02})} / L_{\rm IR}$ ratio  of the $z$-GAL sources is $\sim 10^{-5}$, close to the value found by \citet{Yang2016}. This is consistent with
the expectation that far-IR pumping is likely the 
dominant mechanism of the excitation of the sub-millimetre \hto\  
lines in very dense, heavily obscured 
star-formation-dominated regions \cite[e.g.][]{Gonzalez-Alfonso2014, Yang2016, ga22}. 

The source HeLMS-49 has been detected in the  
ortho-$\rm H_2O(4_{23}-3_{30})$ line ($\nu_\textrm{rest}=448$ GHz), that has an upper energy level of $E_\mathrm{upper}=433$~K. 
This line was first detected in ESO\,320-G030 \citep{PS2017} and later 
in the $z=3.6$ dusty star-forming galaxy G09v1.97 \citep{Yang2020}. 
Both works argue that the origin of the line is mostly far-IR pumping rather than maser emission. Therefore, being optically thin, this highly excited water transition is probing deeply into the dense nuclear regions of these galaxies \citep{ga21}. The luminosity of the $\rm H_2O(4_{23}-3_{30})$ line in HeLMS-49 is $(1.7\pm0.5)\times10^{8}\,L_\odot$, yielding a $L_{\rm H_2O(4_{23}-3_{30})} / L_{\rm IR}$ ratio of $6\times10^{-6}$ (assuming no differential lensing effect). This is about six times larger than what has been found in G09v1.97 \citep[$\sim\,0.9\times10^{-6}$][]{Yang2020}, which might indicate the presence of a strong far-IR source deeply buried in the nuclear region of HeLMS-49, possibly powered by a highly obscured active galactic nucleus (AGN) or a nuclear starburst. Further observations are needed to confirm either of these scenarios.


\section{Molecular gas mass}\label{sect:mgas}

Using the millimetre CO and [C{\small I}] lines detected by NOEMA, as well as the sub-millimetre continuum emitted by dust, in this Section we study the molecular gas reservoir of the $z$-GAL sources, deriving their molecular gas masses and dust to gas mass ratios. 
Star formation takes place in dense molecular clouds and there is little or no correlation between neutral atomic hydrogen and star formation at the low densities \citep[e.g.][see Tacconi et al. \citeyear{tacconi2020} for a review]{kennicutt1989,bigiel2008,bigiel2011}. 
Therefore it is assumed that the fuel of star formation consists predominantly of molecular gas, $M_\textrm{mol}$.

Total and molecular gas masses are defined as:
\begin{eqnarray}
M_\textrm{gas}&=&M_\textrm{mol}+M_\textrm{HI} \textrm{,}\\
M_\textrm{mol}&=&M_{\textrm{H}_2}+M_\textrm{He} = 1.36 \times M_{\textrm{H}_2} \textrm{,}
\end{eqnarray}
where $M_\textrm{He}$ is the mass of helium, $M_{\textrm{H}_2}$ is the mass of molecular hydrogen and 
$M_\textrm{HI}$ the mass of atomic hydrogen. The contribution of helium is factorised as a multiplicative factor $\times1.36$.

The direct detection of $\textrm{H}_2$ is non trivial \citep[see, e.g. the reviews by][]{combes2000,habart2005}: the molecule has no dipole moment and all ro-vibrational and rotational transitions are of quadrupolar origin and faint. Most $\textrm{H}_2$ is in cool, shielded regions with too low excitation to produce bright emission lines and too high extinction to allow a direct detection of UV transitions. Warm $\textrm{H}_2$ mid-IR are virtually invisible at the temperatures of giant molecular clouds (10-20 K) where the bulk of star formation takes place. Finally, near- and mid-IR $\rm H_2$ emission driven by shocks and turbulence traces only a small fraction of the $\textrm{H}_2$ mass of galaxies.

For these reasons, the molecular gas content of galaxies is usually determined using other tracers, in particular carbon monoxide (the most abundant molecule), atomic carbon $[$CI$]$, or more rarely water and HCN, accessible with millimetre spectroscopy. 
In this way, the molecular gas mass is computed from the luminosity of the molecular tracer by:
\begin{equation}\label{eq:alpha}
M_{\textrm{mol}} = \alpha_\textrm{tracer} \, L^\prime_\textrm{tracer}\textrm{,}
\end{equation}
where the conversion factor $\alpha_\textrm{tracer}$ is expressed in units of M$_\odot$ (K km s$^{-1}$ pc$^2$)$^{-1}$.

\subsection{The choice of $\alpha_\textrm{CO}$}\label{sect:alphaCO_choice}

For nearby star-forming galaxies with near-solar metallicity, the commonly adopted $\alpha_\textrm{CO}$ conversion factor is the empirical Milky Way value $\alpha_\textrm{CO,MW}= 4.36\pm 0.9$ M$_\odot$ (K km s$^{-1}$ pc$^2$)$^{-1}$, including the helium contribution \citep[e.g.][]{magnelli2012b,bolatto2013,Carilli-Walter2013,tacconi2020}. For extreme local starbursts (ULIRGs), a long and debated discussion pointing to a $\sim4$ times lower conversion factor has been going on since two decades, initially sustained by evidence from dynamical arguments, and advocating a value of 0.8–1.5 M$_\odot$ (K km s$^{-1}$ pc$^2$)$^{-1}$ for these extreme sources \citep[e.g.][]{downes1998,scoville1997}. A similar value has also been suggested to hold for SMGs, outliers of the main sequence of star formation and powerful DSFGs in general, creating a bimodality between `normal' star-forming galaxies and `starbursts' \citep[e.g.][]{daddi2010,genzel2010,Carilli-Walter2013}. Applying the $1.36\times$ helium correction, this becomes $\alpha_\textrm{CO,SB}=1.09$ M$_\odot$ (K km s$^{-1}$ pc$^2$)$^{-1}$. 

In recent years, this dichotomy has lost its initial popularity and an increasing number of studies, based on (sub-) millimetre spectroscopy and dust continuum observations, highlighted that such a bimodality of the CO conversion factor might in fact be an artificial interpretation of more subtle trends of physical parameters.
\citet{tacconi2020} studied scaling relations of gas content, gas fraction, and depletion timescales as a function of other measurable physical parameters (e.g. stellar mass, $M^\ast$, and specific star formation rate, sSFR) of DSFGs of near-solar metallicity on the MS and above \citep[see also][]{genzel2015,tacconi2018}. 
The depletion timescale of galaxies is considered the primary parameter of gas evolution and these authors point out that the low gas masses inferred from the gas dynamics in local ULIRGs, and more generally for galaxies above the MS, are in this scheme  encapsulated in the dependence of depletion time on the distance from the MS, rather than in a change of $\alpha_\textrm{CO}$ \citep[see also][]{scoville2016,scoville2017}.
\citet{tacconi2020} adopt a metallicity-dependent $\rm ^{12}CO$ conversion factor based on the prescriptions by \citet{bolatto2013} and \citet{genzel2012}, with a reference value $\alpha_\textrm{CO,MW}= 4.36\pm 0.9$ M$_\odot$ (K km s$^{-1}$ pc$^2$)$^{-1}$ at solar metallicity. Gas metallicities are estimated mainly using the stellar mass versus metallicity ($M^\ast-Z$) relation (\citealt{genzel2015}, in the \citealt{pettini2004} scale). Computing gas masses from $\rm ^{12}CO$ detections, dust mass estimates and the sub-millimetre continuum measurements, they show that 
the three tracers are in good agreement, once the conversion zeropoints are correctly cross-calibrated.

Currently the $z$-GAL sample does not include any information about gas metallicity, neither from optical spectroscopy or indirect estimates such as the $M^\ast-Z$ relation. In Sect. \ref{sect:mstar} and Appendix \ref{app:scaling} we develop a method to derive $M^\ast$ from the \citet{tacconi2020} scaling relations, but the resulting $M^\ast$ is defined only modulo a factor that depends on the lens magnification $\mu$. Since the actual value of $\mu$ is currently unknown for the $z$-GAL galaxies, the derived $M^\ast$ are not be adopted in the computation of a metallicity-dependent $\alpha_\textrm{CO}$.

\citet{dunne2022} pointed out that for dusty star-forming galaxies (e.g. selected in the far-IR or sub-millimetre), it is safe to assume that the metallicity is high, such that the gas-to-dust mass ratio, $\delta_\textrm{GDR}$ is broadly similar to that of the MW \citep[see also Appendix \ref{app:scaling} and][]{magdis2012,rowlands2014,berta2016,yang2017}. 
\citet{dunne2022,dunne2021} studied a large, heterogeneous sample of galaxies with observations in three different molecular gas tracers: $^{12}$CO, $[$CI$]$ and sub-millimetre continuum, including main sequence star-forming galaxies up to $z\sim 1$, local ULIRGs, and high-redshift outliers of the MS (thereby labelled SMGs). With an initial choice of the gas mass absorption coefficient $\kappa_\textrm{H}=\delta_\textrm{GDR}/\kappa_{850}=1884$ kg m$^{-2}$, similar to that of the MW and other local disks, these authors retrieve an average $\alpha_\textrm{CO}=4.0\pm0.1$ M$_\odot$ (K km s$^{-1}$ pc$^2$)$^{-1}$, including the contribution of helium (see Tab. 14 of \citealt{dunne2022} for a further review of $\alpha_\textrm{CO}$ values found in the literature).

As a counter-test, \citet{dunne2022} changed their initial assumption into a bimodal normalisation for MS and SMG galaxies and indeed retrieve a final optimised $\alpha_\textrm{CO}$ similarly bimodal. Nevertheless, when this assumption is made, the [C{\small I}] and 850 $\mu$m continuum conversion factors also become bimodal, a fact that is not supported by any observational evidence so far and that would also be a challenge for astro-chemistry models \citep{dunne2022}.
In light of these findings and given the limitations of the available data, in what follows for the $z$-GAL galaxies we adopt a conversion factor $\alpha_\textrm{CO}=4.0$ M$_\odot$ (K km s$^{-1}$ pc$^2$)$^{-1}$, including the contribution of helium.

\subsection{Molecular gas mass from $\rm ^{12}CO$}\label{sect:mgas_CO}

\begin{figure}[!t]   
\centering
\includegraphics[width=0.45\textwidth]{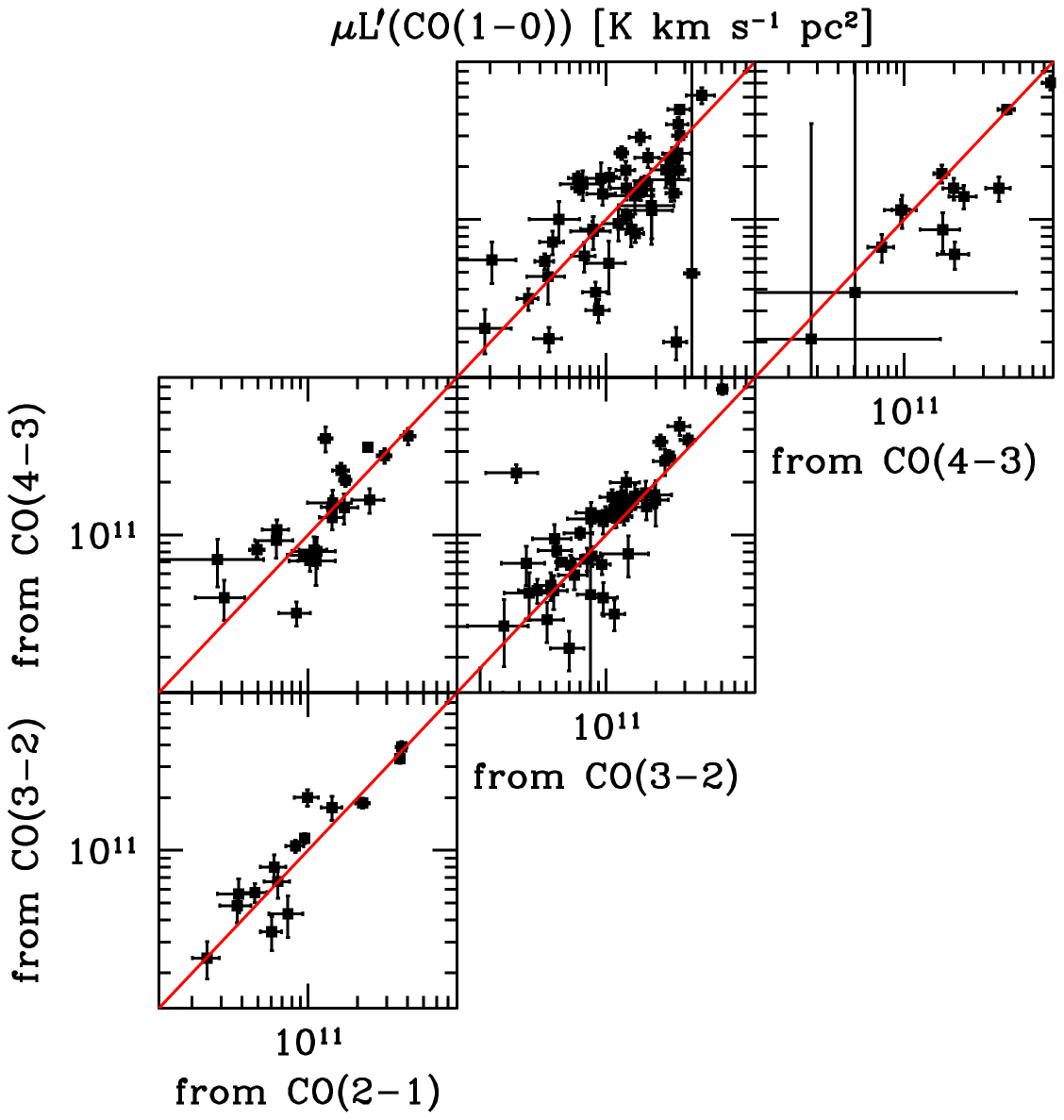}
\rotatebox{-90}{\includegraphics[height=0.45\textwidth]{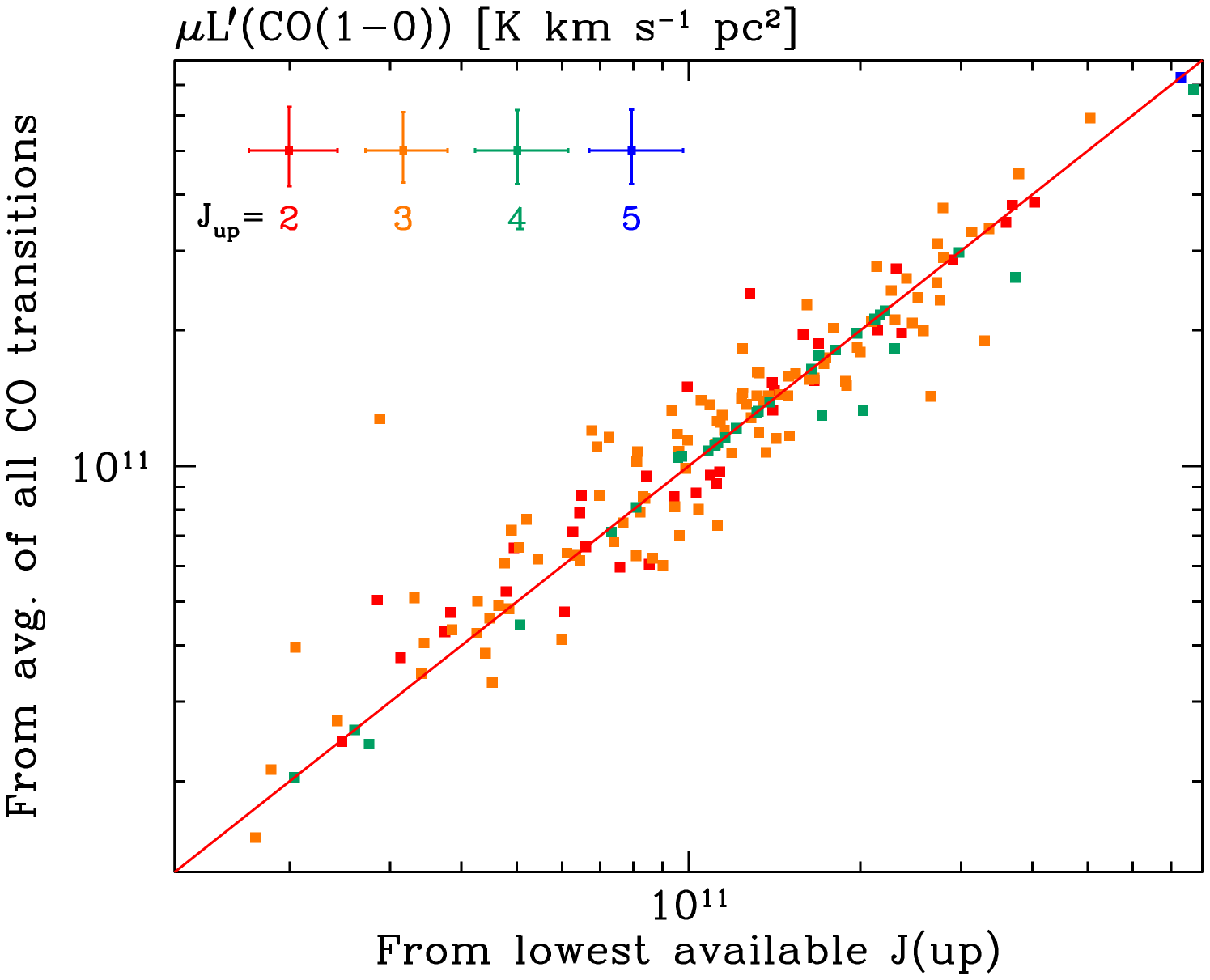}}
\caption{Derivation of $\rm ^{12}CO(1-0)$ luminosity. {\em Upper panels}: Comparison between the values of $L^\prime_\textrm{CO(1-0)}$ derived from $\rm ^{12}CO$ transitions $(2-1)$, $(3-2)$, $(4-3)$ and $(5-4)$ adopting the SMGs line luminosity ratios from \citet{Carilli-Walter2013}. {\em Bottom panel}: Values of $L^\prime_\textrm{CO(1-0)}$ derived from the lowest-J$_\textrm{up}$ $\rm ^{12}CO$ transition available (up to $5-4$), compared to those computed as the average of values from all available $\rm ^{12}CO$ transitions from $(2-1)$ to $(5-4)$. The colour coding specifies which is the lowest $J_\textrm{up}$ available and indicates the typical uncertainties involved. The solid red line shows the 1:1 locus.}
\label{fig:comp_LCO10}
\end{figure}

The luminosity of the $\rm^{12}CO$ lines detected by NOEMA (Sect. \ref{sect:CO_lines}) are transformed into $\rm ^{12}CO(1-0)$ luminosity adopting the average ratios given by \citet{Carilli-Walter2013} for SMGs: $\textrm{r}_{21/10}=0.85$, $\textrm{r}_{32/10}=0.66$, $\textrm{r}_{43/10}=0.46$, and $\textrm{r}_{54/10}=0.39$. For each source, we adopt the $L^\prime_\textrm{CO(1-0)}$ derived from the lowest-$J_\textrm{up}$ detected transition, between $J_\textrm{up}=2$ and 5. Figure \ref{fig:comp_LCO10} compares the values of $L^\prime_\textrm{CO(1-0)}$ derived from the different CO transitions. In the bottom panel, $L^\prime_\textrm{CO(1-0)}$ computed from the lowest available $J_\textrm{up}$ is compared to the average of those computed from all available transitions for the given object up to $(5-4)$, showing a very good agreement.

\begin{figure}[!t]   
\centering
\includegraphics[width=0.45\textwidth]{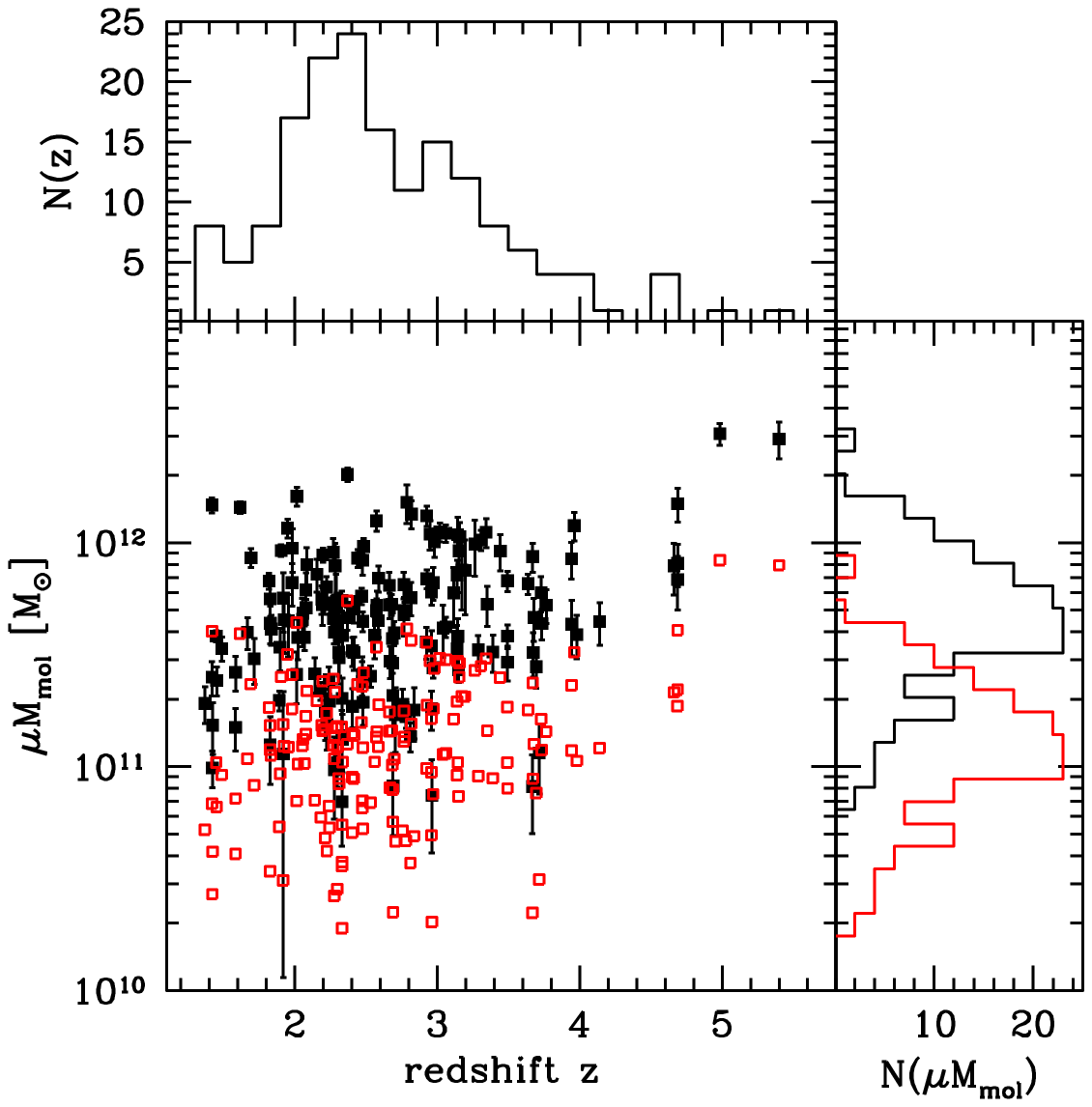}
\caption{Distribution of molecular gas masses of the $z$-GAL sources as a function of redshift, obtained using $\alpha_\textrm{CO}=4.0$ M$_\odot$ (K km s$^{-1}$ pc$^2$)$^{-1}$ (solid black symbols, our choice of preference) and 1.09 (starbursts value, red open symbols), including the contribution of helium.}
\label{fig:distr_Mgas}
\end{figure}

The $\rm ^{12}CO(1-0)$ luminosity derived in this way is finally converted into molecular gas mass, adopting the $\alpha_\textrm{CO}$ value as in Sect. \ref{sect:alphaCO_choice}. Table \ref{tab:Mgas_etc} lists the results and Fig. \ref{fig:distr_Mgas} presents the distribution of $\mu M_\textrm{mol}$ of the $z$-GAL sources. 
The derived values of the molecular gas mass, $\mu M_\textrm{mol}$, not corrected for gravitational magnification, are in the range from $10^{11}$ to few $10^{12}$ M$_\odot$. These lie at the upper end of the $M_\textrm{mol}$ of star-forming galaxies at any redshift, regardless of magnification, comparable to other samples of gravitationally amplified galaxies \citep[e.g.][]{riechers2011, harris2012, ivison2013, yang2017}. See Fig. \ref{fig:KS} for a thorough comparison.

In Fig. \ref{fig:distr_Mgas} we also compare the values of $\mu M_\textrm{mol}$ obtained with our choice of $\alpha_\textrm{CO}$ (black filled symbols) with the one that would have been obtained adopting the starburst normalisation $\alpha_\textrm{CO,SB}=1.09$ M$_\odot$ (K km s$^{-1}$ pc$^2$)$^{-1}$ (red open symbols). As discussed by \citet{dunne2022} a different choice in the normalisation of the conversion factor between the molecular gas tracers and $M_\textrm{mol}$ causes only a simple rescaling of the results.

\subsection{Atomic Carbon}\label{sect:ci}

Out of  the 137 sources in the $z$-GAL (including pilot) sample, 27 sources were detected in the [C{\small I}]($^3$P$_1-^3$P$_0$) emission line, and, for three of them, the [C{\small I}]($^3$P$_2-^3$P$_1$) emission line was also measured. 
The availability of a substantial number of sources where [C{\small I}] is detected allows us to explore the properties of the atomic carbon emission lines and, importantly, to check the consistency of the H$_2$ masses derived independently from the $^{12}$CO and [C{\small I}]($^3$P$_1-^3$P$_0$) emission lines. 
Table~\ref{table:CI-derived-mass} lists the sources where [C{\small I}]($^3$P$_1-^3$P$_0$) and [C{\small I}]($^3$P$_2-^3$P$_1$) are detected  
together with the line fluxes, the [C{\small I}]($^3$P$_1-^3$P$_0$) and ($^3$P$_2-^3$P$_1$) luminosities, the [C{\small I}] masses, the abundance of [C{\small I}] relative to molecular hydrogen, $X[\textrm{C\small{I}}]/X[{\rm H}_2]$, and, in the last column, the molecular gas masses derived from the atomic carbon masses. 

We estimated the neutral carbon masses by using Eq.~1 in \cite{weiss2005}, assuming a [C{\small I}] excitation temperature equal to $T_{\rm exc} = 30 \, K$, which is close to the value in \cite{walter2011}, $<T_{\rm exc}> = 29.1 \pm 6.3$ K, and the mean temperature of $<T_{\rm exc}> = 25.6 \pm 1.0$ K found by \cite{valentino2020}: 
\begin{equation}\label{eq:M-CI}
M_{\rm [C{\small I}]} = 5.706 \times 10^{-4} \, Q(T_{\rm exc}) \frac{1}{3} e^{23.6/T_{\rm exc}} \, L^{\prime}_{\rm [C{\small I}](1-0)} \textrm{,}
\end{equation}
where $Q(T_{\rm exc}) = 1 + 3e^{-23.6 \, {\rm K}/T_{\rm exc}} + 5e^{-62.5 \, {\rm K}/T_{\rm exc}}$ is the partition function of [C{\small I}] and the result is expressed in units of M$_\odot$. 

A more precise measure of the excitation temperature, $\rm T_{exc}$, can be derived for the three sources where the two [C{\small I}] emission lines were detected, namely HeLMS-19, HerBS-185 and HerBS-201. Assuming local thermal equilibrium and, under the condition that the lines are optically thin, the excitation temperature equals the kinetic temperature:
\begin{equation}\label{eq:CI-Texc}
{T_{\rm exc}}/{\rm K} = 38.8/{\ln}(2.11/R_{\rm [C{\small I}]})\textrm{,} 
\end{equation} 
where 
\begin{equation}
R_{\rm [C{\small I}]} = L'_{\rm [C{\small I}](2-1)}/L'_{\rm [C{\small I}](1-0)}\textrm{.}
\end{equation}

Using the measured fluxes of the [C{\small I}]$(2-1)$ emission line, we derive excitation temperatures of 45, 54, and 18~K for HeLMS-19, HerBS-185, and HerBS-201, respectively. A similar scatter in the excitation temperatures has also been found in previous studies \citep[e.g.][]{valentino2020, walter2011, nesvadba2019}, although it has only a minor impact on the estimate of the [C{\small I}] masses.  

Combined with the estimate of $M_\textrm{mol}$ based on $\rm ^{12}CO$, the mass of neutral atomic carbon $M_{\rm [C{\small I}]}$ yields an estimate of the [C{\small I}] abundance relative to molecular hydrogen: $X[{\rm C\small{I}}]/X[{\rm H}_2] = M_{\rm [C{\small I}]}/(6 M_{\rm H_2})$, where $M_{\rm H_2}=M_\textrm{mol}/1.36$. The values of the [C{\small I}] abundance thus obtained are listed in Tab.~\ref{table:CI-derived-mass}; their median value is $1.4\times 10^{-5}$ with a median absolute deviation of $0.5\times 10^{-5}$, consistent with the value of $1.6\times 10^{-5}$ derived by \citet{dunne2022}. 
Therefore the molecular gas masses listed in the last column of Tab.~\ref{table:CI-derived-mass} are derived adopting the average $L^\prime_{\rm [C{\small I}]}/M_\textrm{mol}$ 
conversion factor by \citet{dunne2022}, $\rm  \alpha_{\rm [C{\small I}]} = 17.0 \, M_{\sun} (K \, km s^{-1} \, pc^{-2})^{-1}$, that includes the correction for helium. 

\begin{figure*}[!ht]
\centering
\rotatebox{-90}{\includegraphics[height=0.45\textwidth]{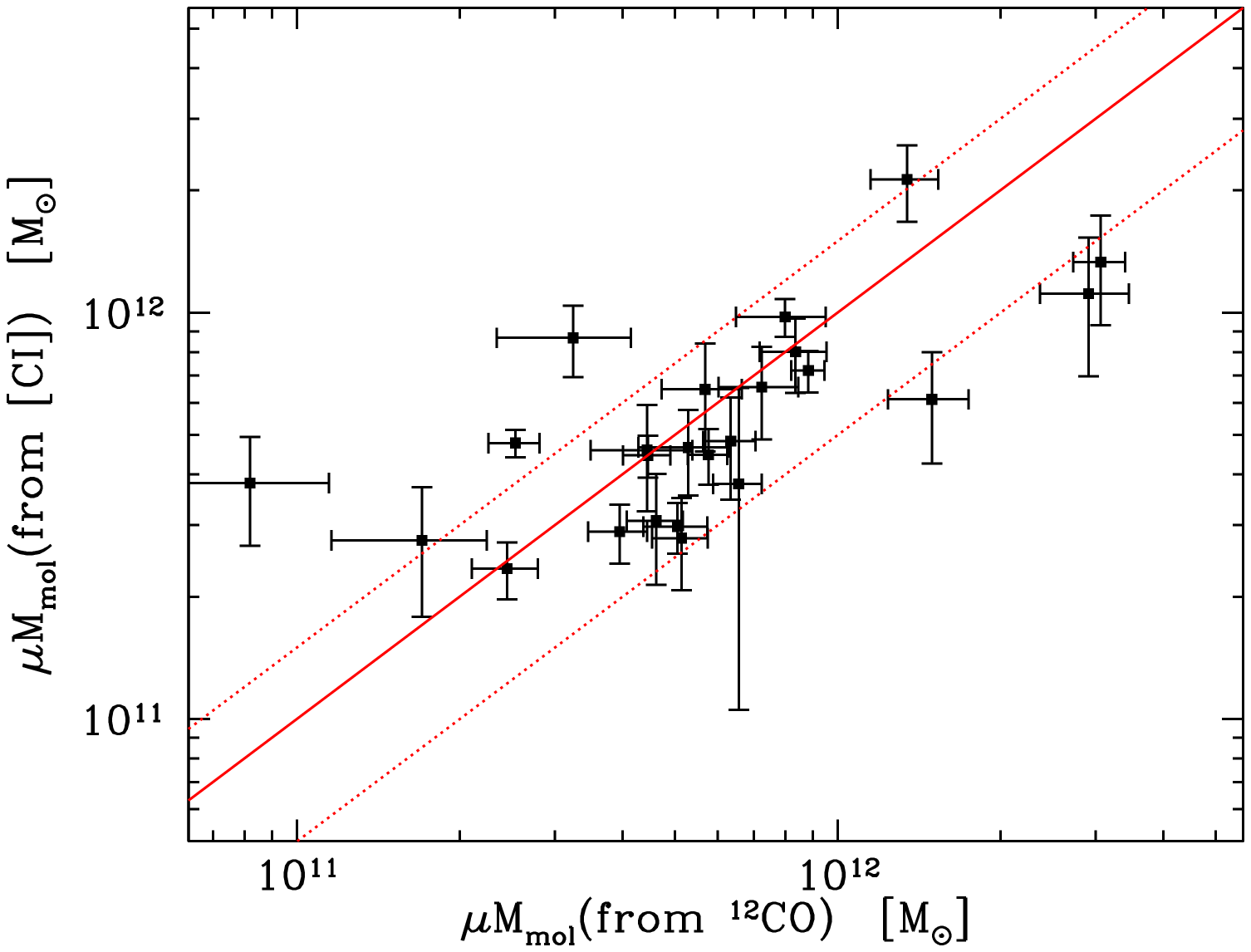}}
\rotatebox{-90}{\includegraphics[height=0.45\textwidth]{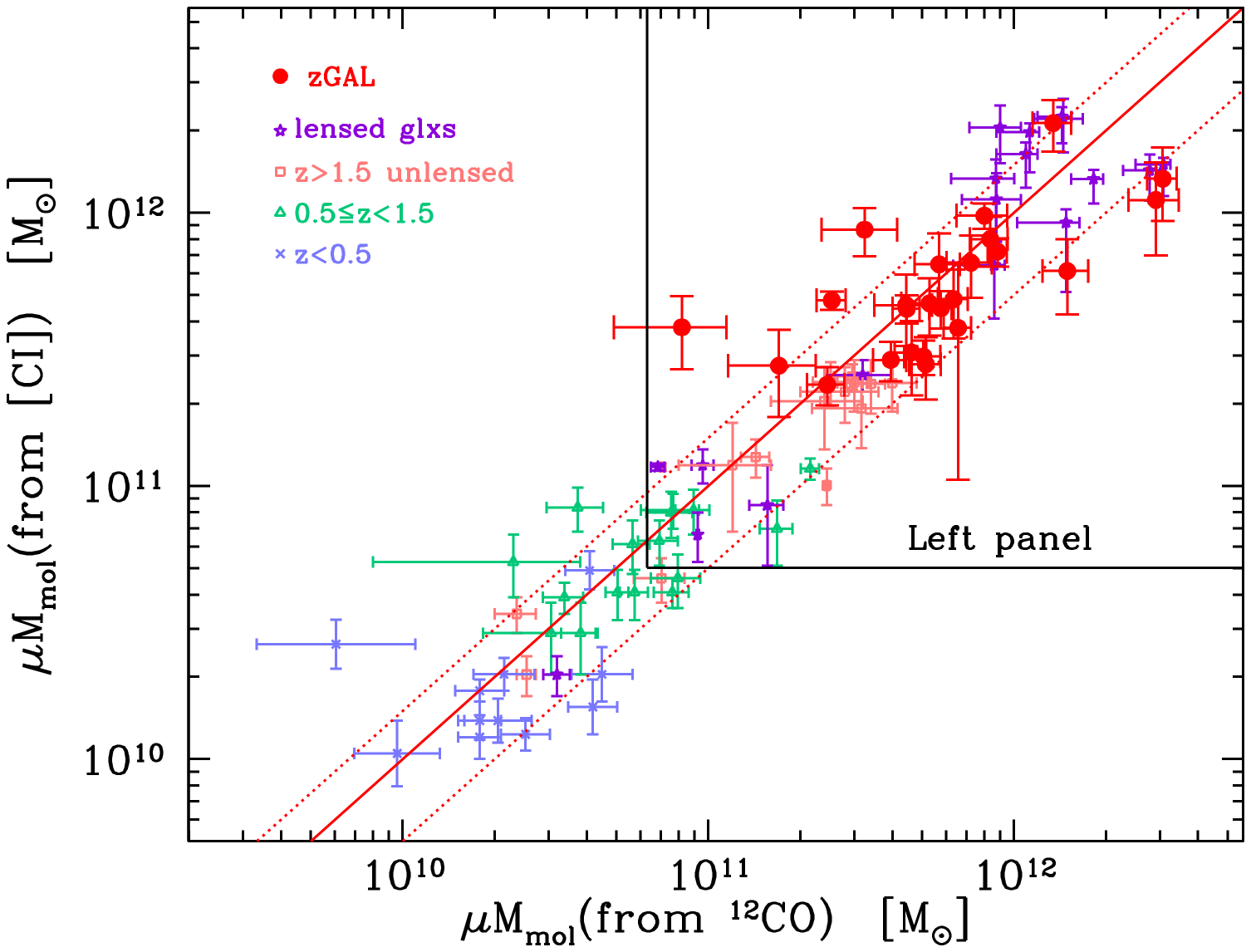}}
\caption{Comparison between the molecular gas masses derived from the [C{\small I}]($^3$P$_1$-$^3$P$_0$) and $\rm ^{12}CO$ emission lines for the $z$-GAL sources where both lines are detected. No correction for amplification was applied to the CO or the [C{\small I}] luminosities. {\em Left panel:} $z$-GAL sources only. The solid red line shows the 1:1 locus and the dotted lines the $\rm \pm 50\%$ region around it. {\em Right panel:} Comparison to sources found in the literature with available [C{\small I}]($^3$P$_1$-$^3$P$_0$) \citep{valentino2018, valentino2020, bothwell2017, dunne2021, alaghbandzadeh2013}. The axes range covered by the left panel is highlighted.}
\label{fig:comp_Mgas_CO_CI}
\end{figure*}

The comparison between the molecular gas mass derived from the lower-$J$ CO emission lines (Sect. \ref{sect:mgas_CO}) and the [C{\small I}](1-0) emission line using the $\rm \alpha_{\rm [C{\small I}]}$ by \citet{dunne2022} is displayed in Fig.\ref{fig:comp_Mgas_CO_CI}. The evidence shown by this diagram is twofold. The consistency between the median [C{\small I}] abundance of the $z$-GAL sample and the value derived by \citet{dunne2022} is reflected by the distribution of the data around the 1:1 locus within a factor of $\sim 1.5$. Moreover, the scatter of the data points in Fig.~\ref{fig:comp_Mgas_CO_CI} (left panel) is the direct consequence of the variety of abundance values that we derived (Tab. \ref{table:CI-derived-mass}).

The {\em right-hand} panel of Fig.~\ref{fig:comp_Mgas_CO_CI} compares the $z$-GAL results to a selection of sources with available [C{\small I}]($^3$P$_1-^3$P$_0$) from the literature. This comparison shows not only the consistency of the adopted $\alpha_\textrm{CO}$ and $\alpha_{\rm [C{\small I}]}$, but also how the $z$-GAL data provide a sizeable addition to previously existing results by populating the locus of lensed sources.

\begin{table*}[!ht]		
\caption{$\rm [C\small{I}]$ Properties of the $z$-GAL sources.}
\label{table:CI-derived-mass}
\centering
\begin{tabular}{lcccccccc}
\hline
\hline
Source name & $z_{\rm spec}$ & $I_{\rm [C\small{I}](1-0)}$ & $I_{\rm [C\small{I}](2-1)}$ & $L'_{\rm [C\small{I}](1-0)}$ & $L'_{\rm [C\small{I}](2-1)}$ & $M_{\rm [C\small{I}]}$ & $X[{\rm C\small{I}}]/X[{\rm H}_2]$ & $M_{\rm mol}(\alpha_{\rm [C\small{I}]})$ \\
 &   & \multicolumn{2}{c}{(Jy km $\rm s^{-1}$)} & \multicolumn{2}{c}{($\rm 10^{10} \, K \, km \, s^{-1} \, pc^2$)}  & $\rm 10^7 \, M_{\sun}$ & $10^{-5}$ & $10^{10} \, M_{\sun}$   \\
\hline
HeLMS-16	& 2.819	& 6.47$\pm$1.38	 & - 		& 12.52$\pm$2.67 & - 		 & 15.52$\pm$3.31 	&   2.6$\pm$0.7	& 212.84$\pm$45.39 \\ 
HeLMS-19 E+W	& 4.688 & 0.78$\pm$0.35 & 2.00$\pm$0.88 & 3.60$\pm$1.10 & 3.41$\pm$1.50 & 4.46$\pm$1.36 	&   0.7$\pm$0.2	& 61.20$\pm$18.7 \\  
HeLMS-20	& 2.195	& 2.05$\pm$0.49 & - 		&  2.74$\pm$0.65 & - 		 & 3.39$\pm$0.80 	&   1.5$\pm$0.4	& 46.58$\pm$11.05 \\
HeLMS-24	& 4.984	& 1.49$\pm$0.45	 & - 		&  7.84$\pm$2.36 & - 		 & 9.72$\pm$2.92  	&   0.7$\pm$0.2	& 133.28$\pm$40.12 \\
HeLMS-26 E	& 2.689 & 1.19$\pm$0.36 & - 		&  2.24$\pm$0.67 & - 		 & 2.77$\pm$0.83 	&   7.7$\pm$3.8	& 38.08$\pm$11.39 \\	
HeLMS-28 	& 2.532	& 1.64$\pm$0.13 & - 		& 2.81$\pm$0.22  & - 		 & 3.48$\pm$0.27 	&   3.1$\pm$0.4	&  47.77$\pm$3.74\\
HeLMS-45	& 5.399	& 1.15$\pm$0.44 & - 		& 6.55$\pm$2.45  & - 		 & 8.12$\pm$3.03 	&   0.6$\pm$0.3	& 111.35$\pm$41.65 \\ 
HeLMS-47	& 2.223	& 2.08$\pm$0.59	 & - 		& 2.84$\pm$0.80  & - 		 & 3.52$\pm$0.99 	&   1.3$\pm$0.4	& 48.28$\pm$13.60 \\      
HeLMS-49	& 2.215	& 1.89$\pm$0.60	 & - 		& 2.56$\pm$0.81  & - 		 & 3.17$\pm$1.00 	&   --		& 43.52$\pm$13.77 \\
HeLMS-51	& 2.156	& 2.99$\pm$0.77	 & - 		& 3.86$\pm$0.99  & - 		 & 4.78$\pm$1.22  	&   1.5$\pm$0.5	& 65.62$\pm$16.83\\
HeLMS-54	& 2.707	& 0.85$\pm$0.30 & - 		& 1.62$\pm$0.57  &  - 		 & 2.01$\pm$0.86 	&   2.7$\pm$1.4	& 27.54$\pm$9.69 \\
HerS-10		& 2.469	& 1.61$\pm$0.25	 & - 		& 2.63$\pm$0.41  &  - 		 & 3.26$\pm$0.50 	&   1.3$\pm$0.2	& 44.71$\pm$6.97 \\
HerS-13		& 2.476	& 2.87$\pm$0.60 & - 		& 4.71$\pm$0.98  &  - 		 & 5.84$\pm$1.21 	&   1.6$\pm$0.4	& 80.07$\pm$16.66 \\
HerS-16		& 2.198	& 3.17$\pm$0.50	 & - 		& 4.24$\pm$0.67  &  - 		 & 5.25$\pm$0.83 	&   1.3$\pm$0.2	& 72.08$\pm$8.50 \\
HerBS-38 SE	& 2.477	& 1.60$\pm$0.19 & - 		& 2.62$\pm$0.31  &  - 		 & 3.25$\pm$0.38 	&   1.7$\pm$0.3	& 44.54$\pm$5.27 \\
HerBS-58	& 2.084	& 4.70$\pm$0.50	 & - 		& 5.74$\pm$0.61  &  - 		 & 7.11$\pm$0.75 	&   2.0$\pm$0.4	& 97.58$\pm$10.37 \\
HerBS-70 E	& 2.307	& 3.50$\pm$0.70	 & - 		& 5.10$\pm$1.02  &  - 		 & 6.32$\pm$1.26  	&   4.4$\pm$1.5	& 86.70$\pm$17.34 \\
HerBS-72	& 3.638	& 0.72$\pm$0.52 & - 		& 2.23$\pm$1.61  &  - 		 & 2.76$\pm$1.99  	&   1.0$\pm$0.7	& 37.91$\pm$27.37 \\
HerBS-85	& 2.817 & 1.86$\pm$0.55 & - 		& 3.81$\pm$1.13  &  - 		 & 4.85$\pm$1.40 	&   1.9$\pm$0.6	& 64.77$\pm$19.21 \\
HerBS-91 C+E	& 2.405 & 1.05$\pm$0.27 & - 		& 1.64$\pm$0.42  &  - 		 & 2.03$\pm$0.52 	&   0.9$\pm$0.3	& 27.88$\pm$7.14 \\
HerBS-115	& 2.370 & 1.19$\pm$0.36 & - 		& 1.81$\pm$0.55  & - 		 & 2.24$\pm$0.68  	&   1.1$\pm$0.4	& 30.77$\pm$9.35 \\
HerBS-143	& 2.240 & 1.00$\pm$0.16 & - 		& 1.38$\pm$0.22  &  - 		 & 1.71$\pm$0.27 	&   1.6$\pm$0.3	& 23.46$\pm$3.74 \\
HerBS-154	& 3.707	& 1.30$\pm$0.40	 & - 		& 4.15$\pm$1.28  & - 		 & 5.14$\pm$1.58  	&    --		& 70.55$\pm$21.76\\
HerBS-169	& 2.698	& 0.89$\pm$0.15	 & - 		& 1.70$\pm$0.28  & - 		 & 2.11$\pm$0.34 	&   1.2$\pm$0.2	& 28.90$\pm$4.76 \\
HerBS-185	& 4.324	& 0.51$\pm$0.34	 & 1.42$\pm$0.30 & 2.07$\pm$1.38  & 0.52$\pm$0.11 & 2.56$\pm$1.71 	&   -- 		& 35.19$\pm$23.46 \\  
HerBS-197	& 2.417	& 1.11$\pm$0.16	 &  - 		& 1.75$\pm$0.25  & - 		 & 2.16$\pm$0.31 	&   1.0$\pm$0.2	& 29.75$\pm$4.25 \\
HerBS-201	& 4.141	& 0.71$\pm$0.21	 & 0.50$\pm$0.10 & 2.70$\pm$0.79  & 0.17$\pm$0.04 & 3.34$\pm$0.98 	&   1.7$\pm$0.6	& 45.90$\pm$13.43 \\
\hline
\end{tabular}
\tablefoot{The $\rm [C{\small I}](1-0)$ and $(2-1)$ line fluxes are from Paper I. Four sources from the Pilot Programme are included, namely: HerBS-58, 70E, 72 and 154 \citep{neri2020}. See text for the derivation of $M_{\rm [C{\rm \small I}]}$ and $M_\textrm{mol}$.}
\end{table*}

\subsection{The 850 $\mu$m continuum}

A common alternative method to derive the molecular gas mass of galaxies is to use their FIR-mm continuum emission to estimate dust mass and then convert it assuming their metallicity properties \citep[e.g.][see \citealt{berta2016} for a discussion]{leroy2011,magdis2012,santini2014,bethermin2015}.
\citet{scoville2013}, \citet{scoville2014,scoville2016}, and \citet{eales2012} proposed that a single frequency, broadband measurement in the Rayleigh-Jeans tail of the dust SED is sufficient to estimate the gas masses. They used nearby objects and distant star-forming galaxies observed with ALMA and Herschel to calibrate the 850 $\mu$m luminosity of a galaxy as a tracer of its gas mass content. 
The basis of the method relies on the fact that the long wavelength Rayleigh–Jeans (RJ) tail of dust emission is predominantly optically thin and can therefore be used to probe 
the total dust mass of the galaxy, if the dust emissivity per unit mass is known.
The gas mass can then be derived assuming a gas-to-dust mass ratio.

Under these conditions and starting from the expression of a MBB (Paper II and \citealt{berta2013}), in the RJ regime the dust emission of a galaxy can be approximated as:
\begin{equation}\label{eq:RJ_lum}
L_\textrm{RJ}\left(\nu\right) = \frac{8\pi k_\textrm{B}}{c^2} \nu_0^2 \kappa_0\left(\frac{\nu}{\nu_0}\right)^{2+\beta} M_\textrm{dust} T_\textrm{dust}\textrm{,}
\end{equation}
where $\kappa_\textrm{dust}\left(\nu\right)=\kappa_0\left(\nu/\nu_0\right)^\beta$ is the dust mass absorption coefficient in units of m$^2$ kg$^{-1}$, $\beta$ is the emissivity index of the dust emission, $M_\textrm{dust}$ the dust mass in kg, $T_\textrm{dust}$ the dust temperature in K, and therefore $L_\textrm{RJ}(\nu)$ is in erg s$^{-1}$ Hz$^{-1}$.

Given the gas-to-dust mass ratio $\delta_\textrm{GDR}$ (Eq. \ref{eq:delta_gdr}), we  define  $\kappa_\textrm{H}=\delta_\textrm{GDR}/\kappa_\textrm{dust}$ in units of kg m$^{-2}$ \citep[e.g.][]{dunne2022}.  Adopting  $\nu_0=\nu_{850}=353$ GHz as reference frequency (corresponding to a wavelength of 850 $\mu$m) to evaluate $\kappa_\textrm{H}=\kappa_{850}$, the conversion factor can be written as: 
\begin{equation}
\alpha_{850} = \frac{L_\textrm{RJ}(\nu)}{M_\textrm{mol}} = \frac{8\pi k_\textrm{B}}{c^2}\nu_{850}^2\frac{1}{\kappa_{850}}\left(\frac{\nu}{\nu_{850}}\right)^{2+\beta}T_\textrm{dust}\textrm{,}
\end{equation}
that enables the computation of $M_\textrm{mol}$, once the 850 $\mu$m rest-frame luminosity of the galaxy is known.

\citet{scoville2014,scoville2016} adopted a fixed dust temperature of 25 K to apply the $k$-correction to the observed flux densities and derive the 850 $\mu$m rest-frame luminosity of their galaxies. They also stressed that the temperature involved in these equations ought to be a mass-weighted temperature, rather than the commonly used luminosity-weighted temperature for example derived with a single-temperature MBB. 
On the other hand, \citet{harrington2021} found that for SMGs a single temperature MBB fit properly reproduces the SED and the dust temperatures thus derived for these objects are consistent with the mass-weighted estimate derived from multi-temperature fits \citep[see also][]{dunne2022}. The consequences of adopting a fixed dust temperature rather than the actual temperature obtained with a dedicated SED fitting for each galaxy are discussed by \citet{dunne2022}. 
\citet{scoville2016} also point out the necessity of applying a further correction $\Gamma_\textrm{RJ}$ to this conversion, to take into account possible deviations from the proper RJ emission law at the wavelength where Eq. \ref{eq:RJ_lum} is evaluated. Similar caveats were also highlighted by \citet{genzel2015}. We defer the reader to all these works for more details on the method and on its application.
More recently, \citet{tacconi2020} and \citet{dunne2022} validated the method and showed how the results are consistent with those of more time consuming spectroscopy, once the relative conversion factors are properly cross-calibrated.

Here we apply the 850 $\mu$m continuum approach to the $z$-GAL galaxies, making full use of the SED fitting presented in Paper II to evaluate the 850 $\mu$m rest-frame luminosity from the best fit models. It is important to note that the $z$-GAL sources benefit from continuum detections that cover wavelengths around and beyond 850 $\mu$m in the rest frame for the majority of the sources and therefore allow for an interpolation between the {\it Herschel}/SPIRE, SCUBA-2, and NOEMA data, rather than relying on extrapolations. Following the success demonstrated for $^{12}$CO and [C{\small I}], we adopt the continuum conversion factor by \citet{dunne2022}: $\alpha_{850}=6.9 \pm 0.1\times10^{12}$ W Hz$^{-1}$ M$_\odot^{-1}$. In their cross-calibration of gas tracers, these authors did not adopt a fixed dust temperature template to apply the $k$-correction, but instead used the results of SED fitting for each galaxy in their sample, similarly to what was done for the $z$-GAL data (Paper II). Table \ref{tab:Mgas_etc} includes the derived values of $\mu M_{\textrm{mol},850\mu\textrm{m}}$ in column 5.

\begin{figure}[!t]
\centering
\rotatebox{-90}{\includegraphics[height=0.47\textwidth]{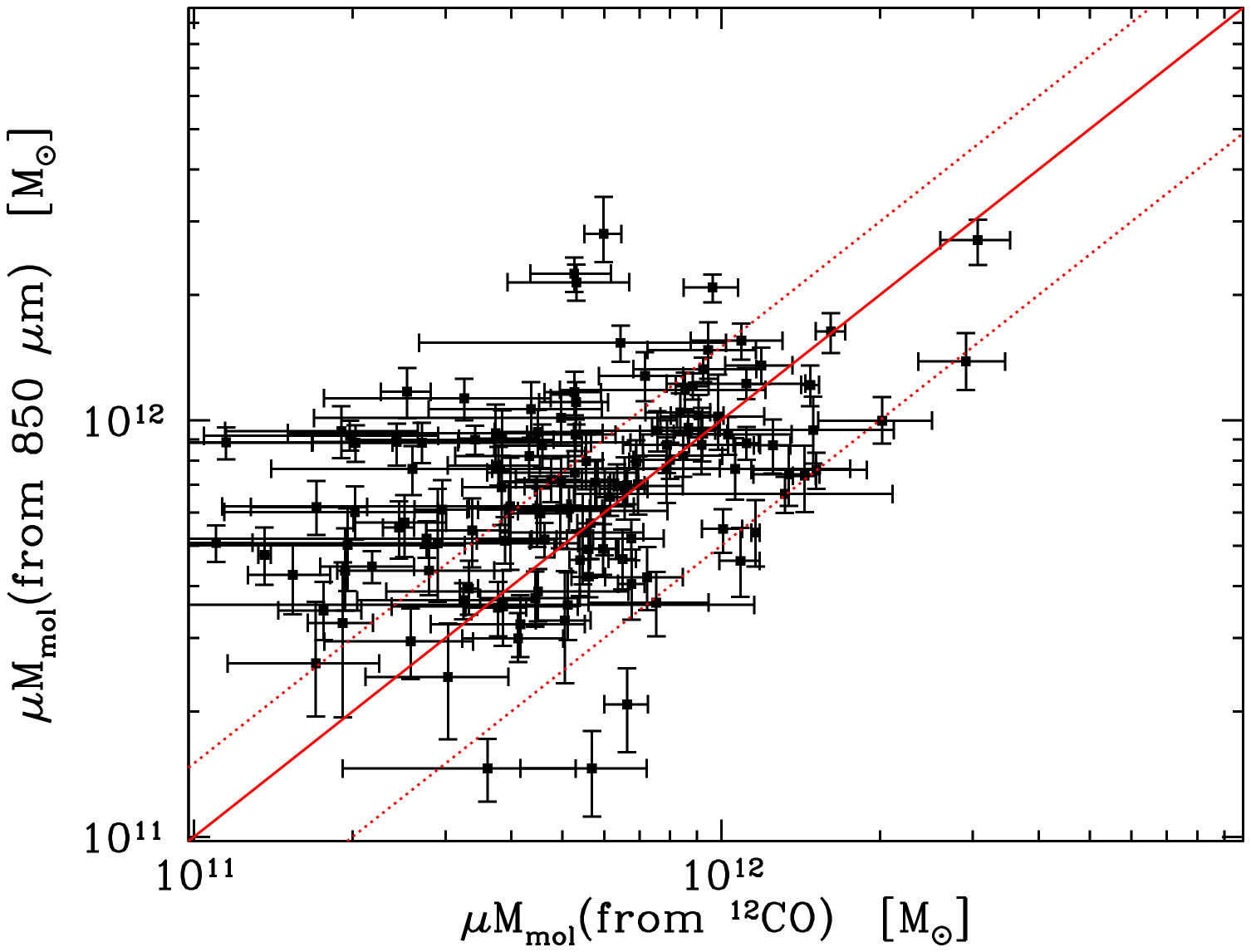}}
\caption{Comparison between the molecular gas masses derived from the 850 $\mu$m continuum and from the $\rm ^{12}CO$ emission lines for the $z$-GAL sources where both tracers are available. No correction for amplification was applied. The solid red line shows the 1:1 locus and the dotted lines the $\rm \pm 50\%$ region around it.}
\label{fig:comp_Mgas_CO_850}
\end{figure}

Figure \ref{fig:comp_Mgas_CO_850} compares the $M_\textrm{mol}$ thus obtained to the one derived from the detected $^{12}$CO transitions. The 850 $\mu$m method produces results consistent to $^{12}$CO within $\pm$50\% for only roughly half of the $z$-GAL sources (i.e. those within the dotted lines in Fig. \ref{fig:comp_Mgas_CO_850}). The median ratio of $M_\textrm{mol}$ from the 850 $\mu$m continuum and from $\rm ^{12} CO$ is 1.23 with a median absolute deviation of 0.51, significantly larger than what is expected from the uncertainty on $\alpha_{850}$ given by \citet{dunne2022}.
We investigated possible reasons for this relatively poor overlap, but we did not find any evident dependencies of the  ratio $M_\textrm{mol}(850\mu\textrm{m})/M_\textrm{mol}(^{12}\textrm{CO})$ on redshift, $L(\textrm{IR})$, line's FWHM or $L^\prime_\textrm{CO(1-0)}$/FWHM. 

The consistency between the gas mass estimates based on $^{12}$CO and [C{\small I}] raises the suspicion that the cause of this large scatter lies in the dust-based method. 
We highlight that we adopted a specific value of $\alpha_{850}$ for our entire sample, 
but the values of $\alpha_{850}$ found in the literature range from 3.6 to 12 \citep[for a review, see Tab 11 of][]{dunne2022}. 
Figure \ref{fig:comp_Mgas_CO_850} might be an indication that the chosen $\alpha_{850}$ conversion factor 
might not be commensurate to the physical properties of all $z$-GAL galaxies. This would somehow not be unexpected, as the sample includes a heterogeneous mix of single lensed sources and interacting pairs, possible main sequence galaxies and starbursts, candidate groups and AGN (Paper IV). Further investigations and higher resolution, deeper observations will be needed to shed more light on each individual object detected by NOEMA.

\subsection{Gas to dust ratio}\label{sect:delta_GDR}

Combining the $M_\textrm{mol}$ derived from the detected $\rm ^{12}CO$ transitions in Sect. \ref{sect:mgas_CO} and $M_\textrm{dust}$ estimated in Paper II through SED fitting, we compute the gas-to-dust ratio of the $z$-GAL sources:
\begin{equation}\label{eq:delta_gdr}
\delta_\textrm{GDR}=\frac{M_\textrm{mol}}{M_\textrm{dust}}\textrm{.}
\end{equation}

Dust masses are based on SED fitting including {\it Herschel}/SPIRE, SCUBA-2 850 $\mu$m (when available) and NOEMA multi-band photometry. In case multiple components with the same redshift were detected by NOEMA for a given {\it Herschel} source, Paper II combined the corresponding millimetre continuum fluxes of the different components. A similar approach is followed here for those sources with multiple detections both in the continuum and lines (at the same redshift). Therefore the number of sources retained amounts to 131.

None of the quantities involved has been corrected for the effects of gravitational lensing magnification. In other words, it is assumed that the continuum dust emission -- on which the derivation of $M_\textrm{dust}$ is based -- and the detected lines defining $M_\textrm{mol}$ are affected by the same magnification and that possible differential effects play a negligible role.

\begin{figure}
\centering
\rotatebox{-90}{\includegraphics[height=0.45\textwidth]{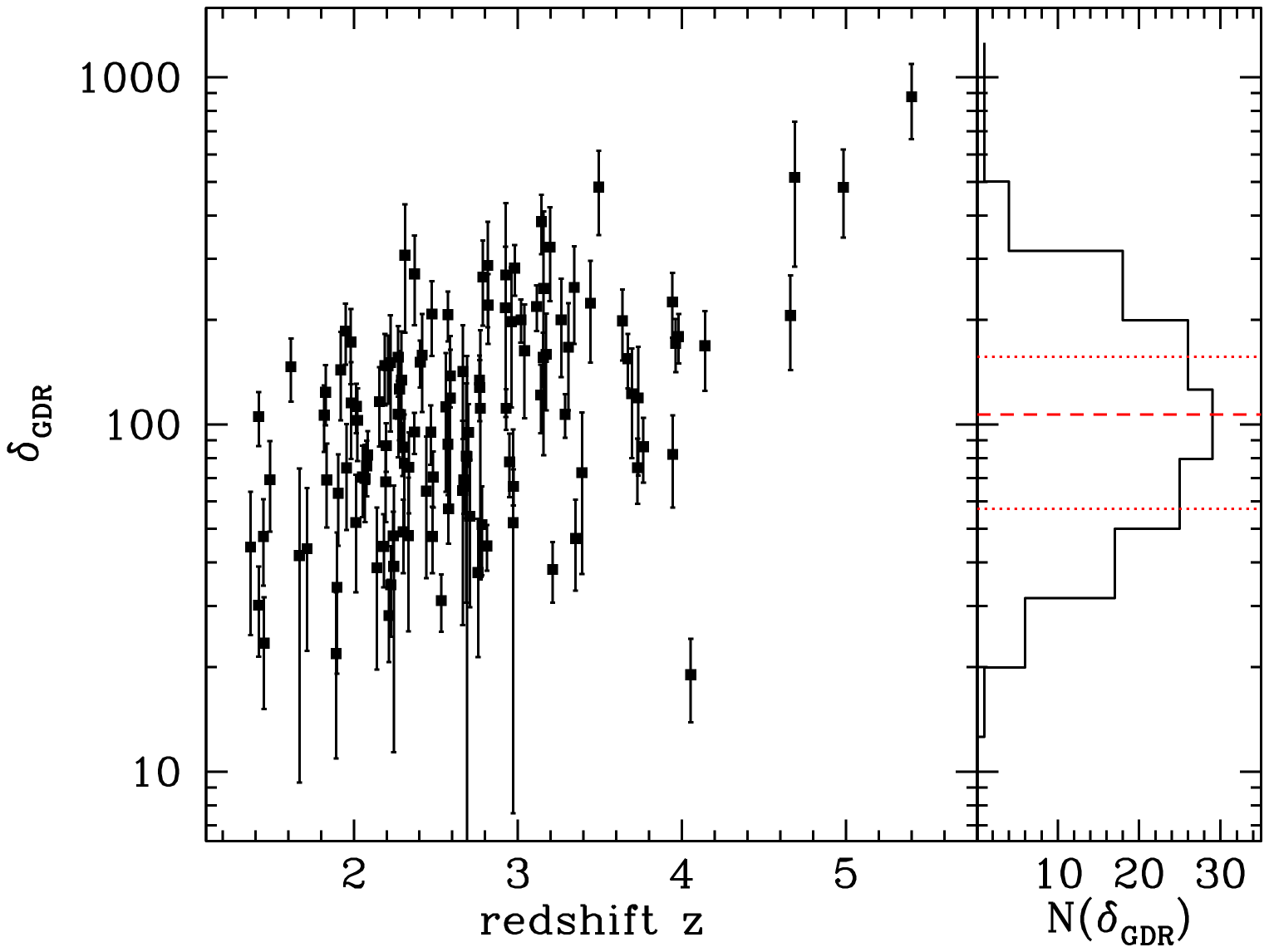}}
\caption{Gas-to-dust mass ratio the $z$-GAL sources. The dashed red line marks the median $\delta_\textrm{GDR}$ of the sample and the two dotted lines indicate its $\pm$ median absolute deviation.}
\label{fig:delta_gdr_test}
\end{figure}

The values of the gas-to-dust ratio of the $z$-GAL sources are listed in Tab. \ref{tab:Mgas_etc}, and Fig. \ref{fig:delta_gdr_test} shows the distribution of $\delta_\textrm{GDR}$ as a function of redshift. The $\delta_\textrm{GDR}$ ratio of the $z$-GAL galaxies covers the range from $\sim20$ to a few 100s, similar to that found by \citet{magdis2012} and \citet{leroy2011}.
The median value of the sample is $\delta_\textrm{GDR}=107$, with a median absolute deviation (m.a.d.) of 50.
For comparison, using $\alpha_\textrm{CO,SB}=1.09$ M$_\odot$ (K km s$^{-1}$ pc$^2$)$^{-1}$, the median value and m.a.d. would become 29.1 and 13.5, respectively.

The actual value of $\delta_\textrm{GDR}$ depends strongly on the underlying assumptions made here, namely the choice of the $\rm ^{12}CO$ conversion factor, which is consistent with MW-like dust and gas properties \citep{dunne2022}, and the values of the dust absorption coefficient $\kappa\left(\nu\right)$ adopted by the MBB SED fitting \citep[][see Paper II]{draine2014}. The derivations of these two quantities having been carried out independently of each other, it is worth to note that the median $\delta_\textrm{GDR}$ of our sample (obtained with our choice of $\alpha_\textrm{CO}=4.0$ M$_\odot$ (K km s$^{-1}$ pc$^2$)$^{-1}$) is consistent with that of main sequence star-forming galaxies of near-solar metallicity \citep[e.g.][]{tacconi2020,magdis2012,leroy2011}.

\section{Integrated Kennicutt-Schmidt relation}\label{sect:KS}

\begin{figure*}
\centering
\rotatebox{-90}{\includegraphics[height=0.8\textwidth]{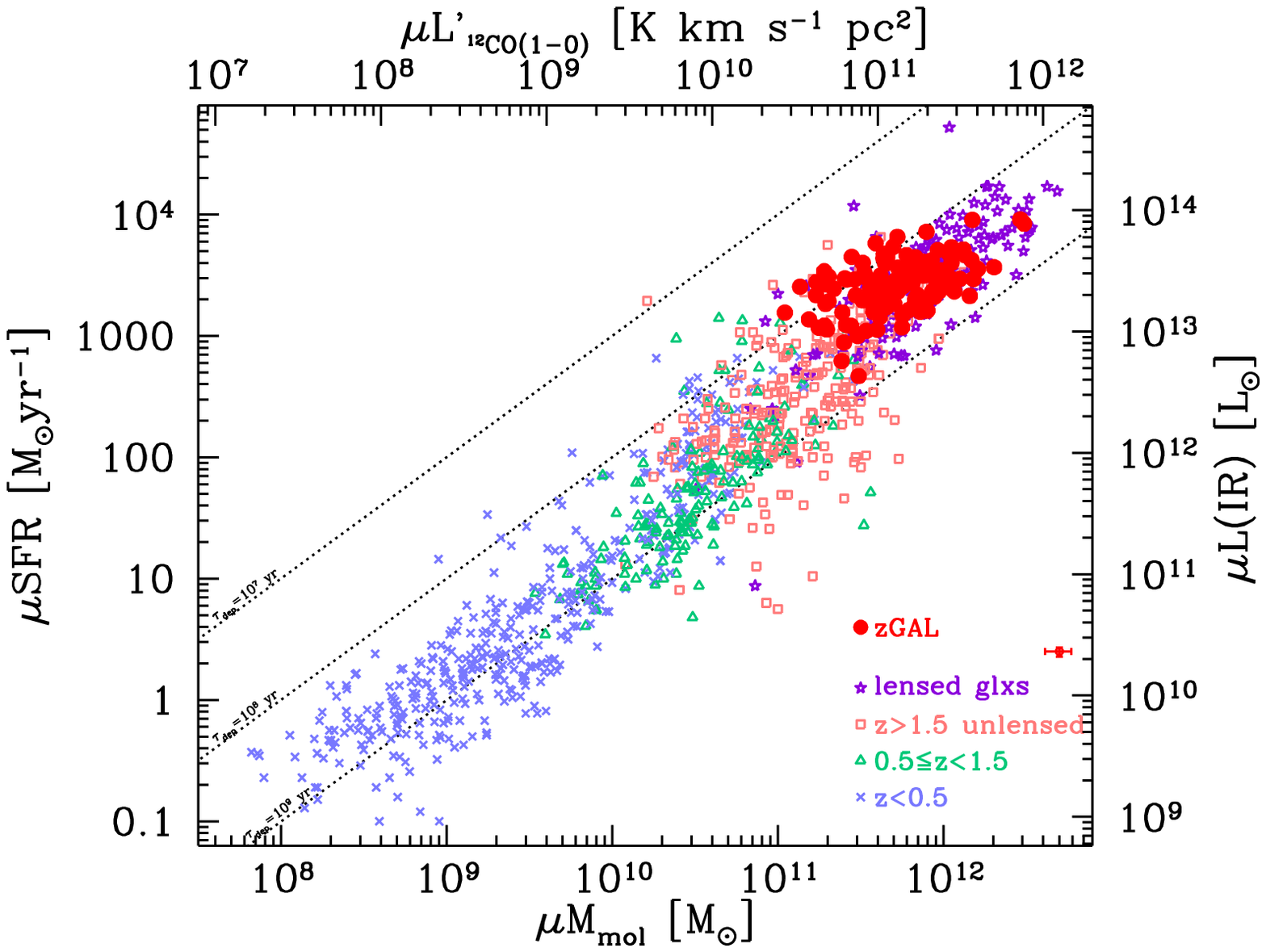}}
\caption{Integrated Kennicutt-Schmidt relation, in terms of the measured luminosities (top and right axes) and of the molecular gas mass and star formation rate (bottom and left axes). The red filled circles represent the $z$-GAL sources, including the Pilot Programme.
The open red symbol and error bar in the bottom-right corner represent the typical (median) $z$-GAL uncertainties.
The dotted lines represent the loci of constant depletion timescales ($\tau_\textrm{dep}=$ 10 Myr, 100 Myr and 1 Gyr). 
The literature data to which the $z$-GAL sources are compared include: at $z<0.5$ the sources studied by \citet{combes2011, combes2013}, \citet{chung2009}, \citet{geach2011}, \citet{solomon1997}, and \citet{villanueva2017}, including local ULIRGs;  
at $z>0.5$, the sources by \citet{alaghbandzadeh2013}, \citet{aravena2016, aravena2014, aravena2013}, \citet{bakx2020}, \citet{bothwell2017, bothwell2013}, \citet{carilli2010}, \citet{dannerbauer2019}, \citet{decarli2016, decarli2019}, \citet{dunne2021, dunne2020b}, \citet{fujimoto2017}, \citet{freundlich2019}, \citet{genzel2015, genzel2003}, \citet{george2013}, \citet{hagimoto2023}, \citet{harris2012, harris2010}, \citet{ivison2013, ivison2011, ivison2010}, \citet{penney2020}, \citet{riechers2020, riechers2011}, \citet{rudnick2017}, \citet{sharon2016}, \citet{tacconi2018, tacconi2013}, \citet{thomson2012}, \citet{valentino2018}, \citet{wang2018}, and \citet{yang2017}.
Lensed objects at $z>1.5$ are marked with empty star symbols.
}
\label{fig:KS}
\end{figure*}

Combining the $^{12}$CO line measurements obtained here and the continuum dust results of Paper II, the relation between the $\rm ^{12}CO(1-0)$ line luminosity and the integrated IR emission is shown in Fig. \ref{fig:KS},
assuming that -- in case of gravitational lensing -- the $^{12}$CO and the FIR-mm continuum emission are co-spatial and similarly magnified. Based on this assumption, accounting for a magnification $\mu$ would shift the affected data points along the diagonal of Fig. \ref{fig:KS}.

This representation of the data is equivalent to the Schmidt-Kennicutt \citep[KS;][]{schmidt1959,kennicutt1998b} relation, integrated over the whole extent of our sources. 
Although it is possible to evaluate the angular extent of the $^{12}$CO and of the sub-millimetre emission of some of these objects (Paper II and Paper IV), it is virtually impossible to retrieve such information for the low resolution {\it Herschel}/SPIRE continuum  measurements. Therefore only the integrated KS space is considered here.

We also compare the $z$-GAL results to a collection of sources found in the literature, split by redshift (see caption of Fig. \ref{fig:KS} for a list of references). 
Lensed sources at $z>1.5$ are marked with empty star symbols. When needed, the molecular gas mass, SFR and other quantities of the sources found in the literature have been recomputed using the same assumptions adopted for the $z$-GAL sample, including the $\alpha_\textrm{CO}$ normalisation, the LIR-to-SFR conversion modified for a Chabrier IMF, and the cosmological parameters.

The $z$-GAL sources sample well the brightest luminosity ($\mu L_\textrm{IR}>10^{13}$ L$_\odot$) and most massive ($M_\textrm{mol}>10^{11}$ M$_\odot$) end of the KS plane, bridging the loci of high-$z$ lensed galaxies and unlensed ULIRGs. Their depletion timescales tend to be on average shorter than those of unlensed samples of more modest luminosity and lower redshift.

Schematically, if a galaxy would consume the entirety of its molecular gas fuel to form stars in one single event at the rate derived from {\it Herschel} and NOEMA data, it would deplete its reservoir in a time given by:
\begin{equation}
\tau_\textrm{dep}=\frac{\mu M_\textrm{mol}}{\mu\textrm{SFR}}\textrm{,}
\end{equation}
called depletion timescale, independent from the magnification $\mu$. The rate of star formation is computed as $\textrm{SFR} = 1.09\times 10^{-10} L_\textrm{IR}$ \citep[][after modification for a \citealt{chabrier2003} IMF]{kennicutt1998a}, with the IR luminosity integrated over the wavelength range $8-1000$ $\mu$m, and expressed in units of L$_\odot$. If $M_\textrm{mol}$ is expressed in units of M$_\odot$ and the SFR in M$_\odot$ yr$^{-1}$, then $\tau_\textrm{dep}$ has naturally units of yr.

\begin{figure*}
\centering
\rotatebox{-90}{\includegraphics[height=0.8\textwidth]{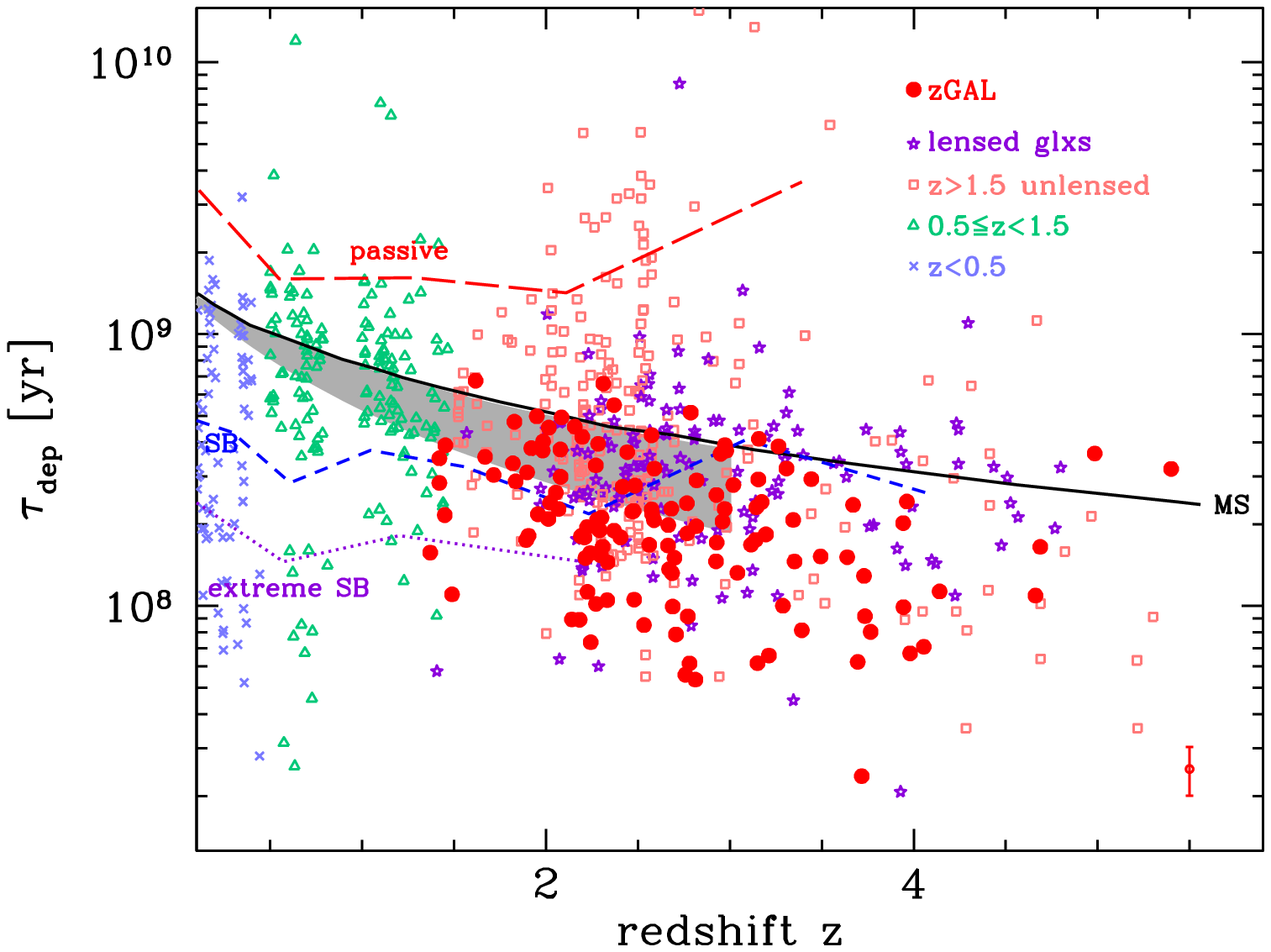}}
\caption{Depletion timescale as a function of redshift. The red filled circles represent the $z$-GAL sources, including the Pilot Programme. The small red symbol and error bar in the bottom-right corner represent the typical uncertainty on $\tau_\textrm{dep}$. The grey shaded area is the trend found by \citet[][]{saintonge2013}. The different lines represent the trends found by \citet{tacconi2020} for  MS galaxies ($\delta_\textrm{MS}=\pm0.6$ dex, black solid line), starburst galaxies ($\delta_\textrm{MS}>0.6$ dex, blue dashed line)), extreme starbursts  ($\delta_\textrm{MS}>1.2$ dex, purple dotted line)), and below-MS galaxies ($\delta_\textrm{MS}<0.4$ dex, red long-dashed line). The collection of data from the literature includes the same data as in Fig. \ref{fig:KS} (see references in the caption).}
\label{fig:taudep_z}
\end{figure*}

The derived values of $\tau_\textrm{dep}$ are reported in Tab. \ref{tab:Mgas_etc} and shown in Fig. \ref{fig:taudep_z}. The evolutionary trends of $\tau_\textrm{dep}$ as a function of redshift are also shown, as determined by \citet{saintonge2013} and \citet{tacconi2020} for MS galaxies (black lines and grey shaded area), outliers of the MS (often associated with starburst galaxies, dashed blue and dotted purple lines), and below-MS galaxies (passive objects, long-dash red line). The majority of the $z$-GAL sources have a depletion timescale in the range between 0.1 and 1.0 Gyr: they occupy part of the main sequence of star formation and the locus of starburst outliers. The tail of the most powerful $z$-GAL sources extends down to $\tau_\textrm{dep}<10^8$ yr, where the most intense bursts of star formation are recorded.

\section{Stellar masses: Inversion of scaling relations}\label{sect:mstar}

The end product of star formation in a galaxy is the mass locked into stars, and is measured by its stellar mass $M^\ast$. The relation between SFR and $M^\ast$ of a galaxy is an indicator of its currently undergoing activity: the position of the source with respect to the so called main sequence of star formation in the $M^\ast-\textrm{SFR}$ plane indicates whether the object is sustained by secular star formation, is undergoing a powerful starburst event, or is a red and dead passive galaxy.

The data currently available for the $z$-GAL sources do not allow us to determine their stellar mass directly. Shorter wavelength observations in the rest-frame near-IR and optical domains would be the ideal tool to this aim \citep[e.g.][]{berta2004}. However, estimating $M^\ast$ is also possible by using the values of $\tau_\textrm{dep}$, as derived in Sect. \ref{sect:KS}.

Scaling relations between the depletion timescale, $\tau_\textrm{dep}$, of a galaxy and its stellar mass content, $M^\ast$, star formation rate, SFR, and position in the $M^\ast$-SFR plane with respect to the main sequence, as a function of redshift, were first derived by \citet{genzel2015}, \citet{scoville2016,scoville2017}, and \citet{tacconi2018} combining observations of the GOODS, COSMOS and other fields that benefit from extensive multi-wavelength coverage.
Combining $^{12}$CO data, {\it Herschel} extragalactic surveys and optical-NIR follow-ups, \citet{tacconi2020} used 2052 star-forming galaxies in the redshift range $0<z<5.3$ to refine these scaling relations, as described by the following equation:
\begin{eqnarray}\label{eq:scaling_maintext}
\log \tau_\textrm{dep} &=& A + B\log\left(1+z\right) + C\log\left(\frac{\textrm{sSFR}}{\textrm{sSFR}\left(\textrm{MS},z,M^\ast\right)}\right)\\
\nonumber &&  +D\left(\log M^\ast -10.7\right)\textrm{.}
\end{eqnarray}
Appendix \ref{app:scaling} describes the best fit parameters $A,\ B,\ C,\ D$ \citep{tacconi2020}. The term $\textrm{sSFR}/\textrm{sSFR}\left(\textrm{MS},z,M^\ast\right)$ represents the distance of a galaxy from the main sequence of star-forming galaxies at the given redshift $z$, in terms of the specific star formation rate $\textrm{sSFR}=\textrm{SFR}/M^\ast$. The adopted MS description is the one derived by \citet{speagle2014}, which has the simple form of a log-linear function of cosmic time $t$ (App. \ref{app:scaling}). We defer to \citet{tacconi2020} for an exhaustive list of the adopted surveys and references. 

For the $z$-GAL sources, the quantities $z$, $\tau_\textrm{dep}$, and SFR have been computed from the NOEMA spectral lines and continuum measurements and from the {\it Herschel} + SCUBA-2 photometry. We can therefore invert this scaling relation (Eq. \ref{eq:scaling_maintext}) to estimate $M^\ast$ for the $z$-GAL galaxies. Appendix \ref{app:scaling} describes this inversion and discusses the possible effect of a metallicity-dependent $\rm ^{12}CO$ conversion factor (see also Sect. \ref{sect:alphaCO_choice}) . Since some of the sources of our sample are gravitationally lensed, this method can provide an estimate of their stellar mass only modulo a factor $\mu^E$, with $\mu$ being the lens magnification and $E=C/\left(C\left(0.84+0.026\, t\right)-D\right)$ a coefficient related to the inversion of Eq. \ref{eq:scaling_maintext} ($t$ is the age of the Universe at the redshift of the source, expressed in Gyr). For the $z$-GAL sample, $E$ is in the range 1.15 (at the high-$z$ end) and 1.28 (at low $z$).

\begin{figure*}
\centering
\begin{minipage}{0.57\textwidth}
\rotatebox{-90}{\includegraphics[height=\textwidth]{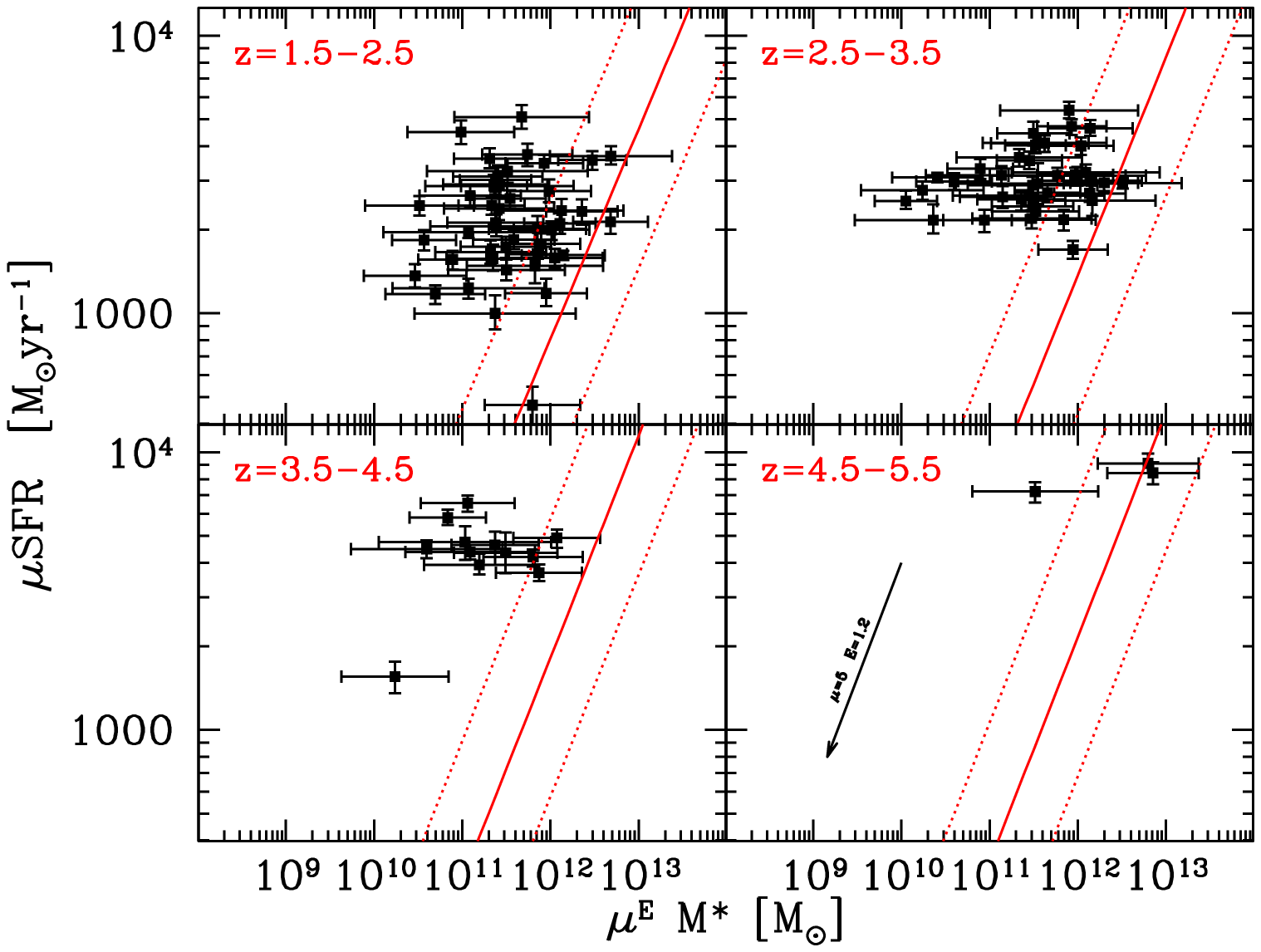}}
\end{minipage}
\begin{minipage}{0.425\textwidth}
\includegraphics[width=\textwidth]{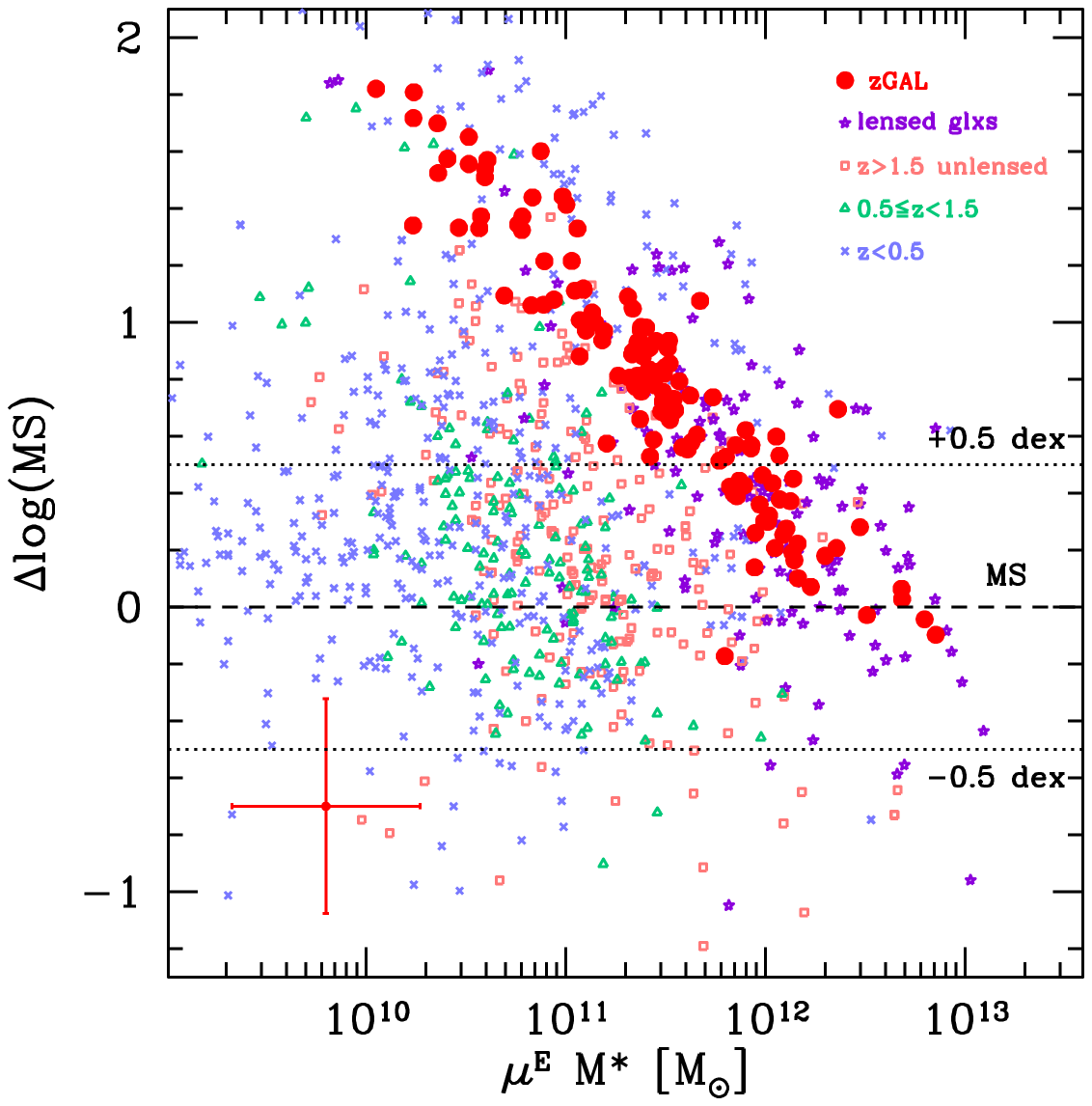}
\end{minipage}
\caption{Analysis of stellar masses, computed by inverting the $\tau_\textrm{dep}$ scaling relation by \citet{tacconi2020} as explained in Appendix \ref{app:scaling}. {\em Left panel:} Position of the $z$-GAL (and pilot) sources in the SFR vs. $M^\ast$ space.  The red solid lines depict the MS of star-forming galaxies \citep{speagle2014} and the dotted lines mark the $\pm0.5$ dex region around the MS. Both SFR and $M^\ast$ are determined modulo the unknown magnification factor $\mu$. The arrow in the lower-right panel indicates the effect of applying a magnification correction with $\mu=5$ and $E=1.2$. {\em Right panel:} Distance from the MS, $\Delta\log(MS)=\log\left(\mu \textrm{SFR}/\textrm{sSFR}\left(\textrm{MS},z,M^\ast\right)\right)$, as a function of stellar mass, $M^\ast$. The dotted horizontal lines represent the range $\pm0.5$ dex from the MS (dashed line). The position of $z$-GAL galaxies is marked with filled circles and literature data are as in Fig.~\ref{fig:KS}. The median uncertainty of the quantities derived for $z$-GAL galaxies is shown in the bottom left. See Sect.~\ref{sect:mstar} and Appendix~\ref{app:scaling} for more details.}
\label{fig:mstar_sfr}
\end{figure*}

A by-product of this analysis is the distance of each $z$-GAL galaxy from the MS: $\Delta\log(MS) = \log \left(\mu \textrm{SFR}/\textrm{sSFR}\left(\textrm{MS},z,M^\ast\right)\right)$, also estimated modulo $\mu$.
The last three columns of Tab. \ref{tab:Mgas_etc} list the results for the three quantities $\mu^E M^\ast$, $E$, and $\Delta \log(MS)$ and Fig. \ref{fig:mstar_sfr} displays the position of the sources in the $M^\ast$-SFR-$z$ space (left) and in the $\Delta\log(MS)$ versus $\mu^E M^\ast$ plane (right). 

The uncertainty on $M^\ast$ is computed via standard error propagation, taking into account the uncertainties in SFR and $\tau_\textrm{dep}$, as well as those on the coefficients of the scaling relation. The resulting large error bars are dominated by the $\tau_\textrm{dep}$ term. 
The arrow in the 4th panel of the $M^\ast$-SFR diagram shows the consequence of applying a magnification correction with $\mu=5$ and $E=1.2$. This correction is basically parallel to the main sequence because the value of $E$ depends on the MS definition (Appendix \ref{app:scaling}).

According to this analysis, the majority of the selected $z$-GAL sources lies above the main sequence of star formation, with only 
15\% and 25\% of them having $\Delta\log(MS)\le \pm 0.3$ or 0.5 dex, respectively, in line with the analysis of the depletion times scales as a function of redshift shown in Fig. \ref{fig:taudep_z}. The stellar masses $\mu M^\ast$ derived with this method are distributed in the range from few $10^{10}$ to over $10^{12}$ M$_\odot$. These extreme values strongly depend on the possible gravitational lensing amplification $\mu$ and on the adopted parametrisation of the MS (Eqs. \ref{eq:scaling_maintext} and \ref{eq:MS}).

The distribution of sources as a function of distance from the MS and of $M^\ast$ (right-hand diagram in Fig. \ref{fig:mstar_sfr}) reveals a strong selection effect for the $z$-GAL galaxies. Because of the very bright {\it Herschel} 500 $\mu$m flux cut and the peaked redshift distribution, our sources occupy a tight locus that consists of powerful starbursts at low masses and intersects the MS only at the high mass end. Arguably, the largest lensing magnification corrections occur for galaxies with the largest apparent stellar masses.

High resolution, deeper observations are required to estimate the value of $\mu$ for each source, by means of lens modelling. Future optical-NIR data, also at high angular resolution (in particular with {\it JWST}), will sample the stellar emission of these sources and will deblend them from the possible foreground lens. At the same time, optical-NIR spectroscopy will help to constrain also the metallicity of these sources.  The combination of these three pieces of information will shed light on the actual stellar mass of the $z$-GAL sources and will be the final test for the scaling inversion method pioneered here.

\section{Summary and concluding remarks}
\label{sect:conclusions}

In this paper we presented a detailed analysis of the $z$-GAL sample, based on NOEMA millimetre observations of both dust continuum and lines emission, and comprising 165 individual sources with robust spectroscopic redshift (Paper I and \citealt{neri2020}). The detected spectral emission lines include $\rm ^{12}CO$, \hto, and [C{\small I}] transitions; 81\% of the sources have two $\rm ^{12}CO$ lines detected, while 8\% have three. 
By combining the spectral information with the continuum results presented in Paper II, we derived the properties of the molecular gas of the sample and constrained the physical nature of these sources. The main findings of this study are as follows:

\begin{itemize}
\item The $\rm ^{12}CO$ line luminosity ratios of the $z$-GAL sample and their average SLED normalised by $L_\textrm{IR}$ have shown that the $\rm ^{12}CO$ ladder on average resembles that of high-redshift SMGs and of the local ULIRG Arp 220, that are powered by bursty star formation \citep{Carilli-Walter2013,yang2017,rangwala2011}. The analysis of the SLEDs of seven individual $z$-GAL sources with three $\rm ^{12}CO$ transitions detected by NOEMA shows that the molecular gas temperature is in the range 100-250 K and that its density is between $10^{3.3}$ and $10^{4.0}$ cm$^{-3}$, consistent with the values found in other high-redshift SMGs \citep{yang2017, canameras2018, harrington2021, stanley2022}.
\item In seven $z$-GAL sources the para-$\rm H_2O(2_{11}-2_{02})$ transition has been detected. 
Including these sources in the water versus $L_\textrm{IR}$ correlation, we derive $L_{\rm H_2O(2_{11}-2_{02})} \sim L_{\rm IR}^{0.97\pm0.08}$, slightly shallower than previous findings \citep[e.g.][]{Yang2016}, but in agreement within the uncertainties of the slope.
\item The molecular gas mass of the $z$-GAL sources has been computed from their estimated $\rm ^{12}CO(1-0)$ luminosities, using a conversion factor 
$\alpha_\textrm{CO}=4.0$ M$_\odot$ (K km s$^{-1}$ pc$^2$)$^{-1}$ \citep[following][]{dunne2022}. The available $\rm ^{12}CO$ transitions have been translated to the $(1-0)$ transition by adopting typical line luminosity ratios of SMGs \citep{Carilli-Walter2013}. The different $\rm ^{12}CO$ transitions available in $z$-GAL produce similar results. The derived masses cover the range $\mu M_\textrm{mol}=10^{11}$ to few $10^{12}$ M$_\odot$.   
\item In parallel, molecular gas masses have also been computed from the available [C{\small I}] transitions and from the 850 $\mu$m rest-frame continuum. The former produces an estimate of $M_\textrm{mol}$ consistent with $\rm ^{12}CO$ within a factor $\sim1.5$. The latter gives results consistent with $\rm ^{12}CO$ within $\pm50$\% for only half of the $z$-GAL sample. This result is not unexpected: the $z$-GAL selection is a simple 500 $\mu$m flux cut combined with a photometric redshift restriction (Paper I) and produces a rather heterogeneous sample in terms of physical properties. Therefore the assumptions of a single value of $\alpha_{850}$ might not be appropriate for all $z$-GAL galaxies, considering that the range of values found in the literature is as large as a factor of three \citep{dunne2022}.
\item Combining the information derived from the $\rm ^{12}CO$ spectra and the NOEMA continuum (Paper II), we derived the gas-to-dust mass ratio, $\delta_\textrm{GDR}$, of the $z$-GAL sources. The median value is 107, with a median absolute deviation of 50, consistent with the values of star-forming galaxies of nearly solar metallicity.
\item The same combined analysis of millimetre spectral lines and dust continuum allowed us to build the integrated Kennicutt-Schmidt relation \citep{schmidt1959,kennicutt1998b} linking the ongoing rate of star formation of the galaxies to their molecular gas reservoir. The ratio of these two quantities is the depletion timescale of the available $M_\textrm{mol}$ at the current SFR and is independent of possible lensing magnification (assuming no differential effects between the dust continuum and the CO emitting regions). The $z$-GAL sources were found to have $\tau_\textrm{dep}$ in the range between 0.1 and 1.0 Gyr, located between the main sequence, where secular star formation takes place, and the above-MS outliers, where starburst events dominate.
\item Finally, inverting the scaling relations defined by \citet{genzel2015} and \citet{tacconi2020} that link depletion timescales to other physical properties of star-forming galaxies, we estimated the stellar mass content of the $z$-GAL sources, modulo the possible magnification due to gravitational lensing. 
The results of this new method confirm that the $z$-GAL sample is mostly composed of sources above the main sequence, powered by strong star formation activity.
\end{itemize}

Despite the heterogeneity of the $z$-GAL sample, coming from a simple {\it Herschel}/SPIRE flux cut and resulting in a variety of sources including possible proto-cluster members, confirmed AGNs, lensed galaxies, multiple systems, interacting pairs, and isolated HyLIRGs (Papers I, II, IV, and \citealt{stanley2022,berta2021,neri2020}), the common denominator of the $z$-GAL survey is that the majority of our objects seem to host powerful starbursts destined to exhaust their molecular gas reservoir over timescales of the order of few $10^8$ years.

In addition to the starbursts population, a fraction of the observed objects belong to the main sequence of star formation, where galaxies undergo a more secular evolution rather than being in a short lived starbursting phase. Roughly 25\% of the $z$-GAL sources lie within $\Delta\log(MS)<\pm0.5$ dex and $\sim15$\% within 0.3 dex.

The broad band capability of NOEMA and its high spectral sensitivity, enabled by the Polyfix correlator, have uncovered the $z$-GAL treasure trove and have made possible -- along with {\it Herschel} and SCUBA-2 data and VLA follow-up observations -- to unveil the star formation properties of these galaxies.  
Deeper and higher-resolution multi-wavelength observations are now required to study these sources in greater detail. Optical and near-IR data will directly constrain the stellar component, thus corroborating, refining or disproving the estimates of $M^\ast$ presented here. Higher resolution sub-millimetre data will enable to verify the lensed nature of many of these galaxies and reconstruct their structure in the source plane via gravitational lens modelling (e.g. \citealt{berta2021}; Borsato et al., submitted to MNRAS). Mid-IR spectroscopy (e.g. with the {\it James Webb Space Telescope}) will shed light on feedback and gas accretion mechanisms, probe the dissipation of kinetic energy by turbulence, and unveil the properties of the hot molecular gas and warm dust components of the ISM in these galaxies. Finally, X-ray and high-energy data (e.g. coming from the all-sky {\it e-Rosita} survey), combined with JVLA follow-up, will ultimately characterise the active galactic nuclei identified in some of the $z$-GAL galaxies. 
The many secrets of the $z$-GAL treasure trove are still to be unveiled.


\begin{acknowledgements}
We are thankful to the anonymous referee for their useful suggestions that helped to improve the content of this paper, and to L.~Tacconi and R.~Genzel for insightful discussions about scaling relations. 
We recognise the essential work of the $z$-GAL Cat Team and Tiger Team, who performed the calibration, reduction and delivery of the $z$-GAL data.
The authors would like to thank I. Cortzen and C. Herrera for their contribution in the early stages of the project and wish them success in the new career. SB thanks GN'R for accompanying him during the intense $z$-GAL analysis times. We also highlight the inspiring role of Jacques de Chabannes, seigneur de La Palisse, and of Jonathan Livingston Seagull. 
This work is based on observations carried out under project numbers M18AB and subsequently D20AB, with the IRAM NOEMA Interferometer. IRAM is supported by INSU/CNRS (France), MPG (Germany) and IGN (Spain). The authors are grateful to IRAM for making this work possible and for the continuous support that they received over the past four years to make this large programme a success. The authors are also grateful to the IRAM director for approving the DDT proposal that enabled to complete the survey. 
This work benefited from the support of the project $z$-GAL ANR-AAPG2019 of the French National Research Agency (ANR). 
AJB and AJY acknowledge support from the National Science Foundation grant AST-1716585.
AN acknowledges support from the Narodowe Centrum Nauki (UMO-2020/38/E/ST9/00077).
CY acknowledges support from ERC Advanced Grant 789410.
DAR acknowledges support from the National Science Foundation under grant numbers AST-1614213 and AST-1910107 and from the Alexander von Humboldt Foundation through a Humboldt Research Fellowship for Experienced Researchers as well as from the Deutsche Forschungsgemeinschaft (DFG) through SFB~956. 
HD acknowledges financial support from the Agencia Estatal de Investigaci{\'o}n del Ministerio de Ciencia e Innovaci{\'o}n (AEI-MCINN) under grant (La evoluci{\'o}n de los c{\'i}umulos de galaxias desde el amanecer hasta el mediod{\'i}a c{\'o}smico) with reference (PID2019-105776GB-I00 / DOI:10.13039/501100011033) and acknowledges support from the ACIISI, Consejer{\'i}a de Econom{\'i}a, Conocimiento y Empleo del Gobierno de Canarias and the European Regional Development Fund (ERDF) under grant with reference PROID2020010107.
RJI acknowledges funding by the Deutsche Forschungsgemeinschaft (DFG, German Research Foundation) under Germany's Excellence Strategy -- EXC-2094 -- 390783311.
SJ is supported by the European Union's Horizon Europe research and innovation programme under the Marie Sk\l{}odowska-Curie grant agreement No. 101060888.
SS was partly supported by the ESCAPE project. ESCAPE - The European Science Cluster of Astronomy \& Particle Physics ESFRI Research Infrastructures has received funding from the European Union's Horizon 2020 research and innovation programme under Grant Agreement no. 824064. 
TJLCB acknowledges support from NAOJ ALMA Scientific Research Grant Nos. 2018-09B and JSPS KAKENHI No.~17H06130, 22H04939, and 22J21948.
\end{acknowledgements}

\bibliographystyle{aa} 		
\bibliography{zgal_mgas} 	



\onecolumn
\begin{appendix}

\section{Inversion of the $\tau_\textrm{dep}$ scaling relation in the case of lensed galaxies}\label{app:scaling}

\citet{genzel2015} defined scaling relations linking the depletion timescale and the molecular gas fraction of galaxies to their other fundamental physical properties, redshift $z$, stellar mass $M^\ast$, and distance from the main sequence in terms of specific star formation rate $\textrm{sSFR}/\textrm{sSFR}(MS,z,M^\ast)$ \citep[see also][]{scoville2016,scoville2017,tacconi2018}.
\citet{tacconi2020} refined these scaling relations using 2052 unlensed star-forming galaxies $z=0$ and $z=5.3$, with stellar masses in the range $\log M^\ast=9.0-12.2$, and distances from the MS between -2.6 and +2.2 dex. The latter parameter indicates that the study has included not only MS galaxies, but also objects below the MS (also considered as `passive galaxies') as well as outliers above the MS (i.e. starbursts). 

By means of variables separation, \citet{tacconi2020} parametrise the depletion timescale of a galaxy as:
\begin{equation}\label{eq:scaling}
\log \tau_\textrm{dep} = A + B\log\left(1+z\right) + C\log\left(\frac{\textrm{sSFR}}{\textrm{sSFR}\left(MS,z,M^\ast\right)}\right)  +D\left(\log M^\ast -10.7\right)\textrm{,}
\end{equation}
with $A=+0.21\pm0.1$, $B=-0.98\pm0.1$, $C=-0.49\pm0.03$ and $D=+0.03\pm0.04$. In this computation, $\tau_\textrm{dep}$ is expressed in units of Gyr and the other quantities in solar units.
The parametrisation of the MS adopted in this study is the one defined by \citet{speagle2014}:
\begin{equation}\label{eq:MS}
\log \textrm{SFR}\left(\textrm{MS},M^\ast,t\right) = \left(0.84-0.026\, t\right)\log M^\ast - \left(6.51-0.11 t\right)\textrm{,}
\end{equation}
where we have omitted the uncertainties on the parameters for simplicity's sake. In this expression, $t$ is the age of the Universe in Gyr at the redshift $z$. \citet{speagle2014} also provide a parametrisation of the MS as a function of redshift instead of $t$, but they claim that it is less accurate, in addition of being a more complex mathematical expression. The age of the Universe at a given redshift is the difference between its current age $t_0$ and the look back time from today to that redshift:
\begin{equation}
t\left(z\right)=t_0 - t_\textrm{H}\int_{0}^{z}\frac{dz^\prime}{\left(1+z^\prime\right)\sqrt{\Omega_m\left(1+z^\prime\right)^3+\Omega_\Lambda}}\textrm{,}
\end{equation}
for the cosmology adopted here, and $t_\textrm{H}$ being the Hubble time.

The question is if, given the data in hand, it is possible to invert the \citet{tacconi2020} scaling relation to derive a first, rough estimate of $M^\ast$. Expliciting Eq. \ref{eq:scaling}, we obtain:
\begin{eqnarray}
\nonumber \log \tau_\textrm{dep} &=& A + B\log\left(1+z\right) + C\log\left(\frac{\textrm{SFR}}{M^\ast}\frac{M^\ast}{\textrm{SFR}\left(\textrm{MS},z,M^\ast\right)}\right) +D\left(\log M^\ast -10.7\right)\\
\nonumber &=& A + B\log\left(1+z\right) + C\log\left(\frac{\textrm{SFR}}{\textrm{SFR}\left(\textrm{MS},z,M^\ast\right)}\right)  +D\left(\log M^\ast -10.7\right)\textrm{.}
\end{eqnarray}
Inserting Eq. \ref{eq:MS} we finally get:
\begin{eqnarray}
\nonumber \log \tau_\textrm{dep} &=& A + B\log\left(1+z\right) + C \log \textrm{SFR} - C \left( \left(0.84-0.026\, t\right)\log M^\ast -\left(6.51-0.11\, t\right)\right) +D\log M^\ast - 10.7 D \\
\nonumber &=& \left(A-10.7D\right) +B\log\left(1+z\right)+C\log\textrm{SFR} -\left(C\left(0.84-0.026\, t\right)-D\right)\log M^\ast +C\left(6.51-0.11t\right)\textrm{.}
\end{eqnarray}
Knowing the values of $\tau_\textrm{dep}$, $z$, $t(z)$ and SFR, we obtain an estimate of $M^\ast$ by inverting the last equation:
\begin{equation}\label{eq:Mstar_no_lensing}
\log M^\ast = \frac{\left(A-10.7D\right)+B\log\left(1+z\right)+C\log\textrm{SFR} +C\left(6.51-0.11\, t\right)-\log\tau_\textrm{dep}}{C\left(0.84-0.026\, t\right)-D}\textrm{.}
\end{equation}

Some $z$-GAL sources are gravitationally lensed (Paper IV and \citealt{berta2021}) and SFR is only known modulo their magnification $\mu$. In such a case, Eq. \ref{eq:Mstar_no_lensing} needs to be adapted: the multiplicative factor $\mu$ enters in the $\log\textrm{SFR}$ term and consequently an equivalent term must be added in the $M^\ast$ side of the equation:
\begin{equation}\label{eq:Mstar_yes_lensing}
\log M^\ast + \frac{C\log\mu}{C\left(0.84+0.026\, t\right)-D} = \frac{\left(A-10.7D\right)+B\log\left(1+z\right)+C\log\left(\mu\textrm{SFR}\right) +C\left(6.51-0.11\, t\right)-\log\tau_\textrm{dep}}{C\left(0.84-0.026\, t\right)-D}\textrm{.}
\end{equation}
We define the extra term $E=C/\left(C\left(0.84+0.026\, t\right)-D\right)$, which for the $z$-GAL sources has values in the range 1.15 (at the high-$z$ end of the sample) and 1.28 (at low $z$). Thus this method allows us to estimate the quantity $\mu^EM^\ast$ only, until a measurement of magnification will be available.

\subsection{The effect of metallicity}

As mentioned in Sect. \ref{sect:alphaCO_choice}, \citet{tacconi2020} adopted a metallicity-dependent $\rm ^{12} CO$ conversion factor, but the $z$-GAL data do not include any information about the sample's metallicity. In the case of dusty star-forming galaxies, such as the $z$-GAL sources, selected with a bright far-IR flux cut, metallicity is expected to be high and not too dissimilar from the MW value.
\citet{berta2016} and \citet{magdis2012} studied the metallicity and $\delta_\textrm{GDR}$ of high-$z$ star-forming galaxies detected by {\it Herschel}, including MS and BzK galaxies, SMGs and lensed sources. The metallicities of these sources were shown to be nearly solar or slightly sub-solar \citep[down to 0.5~Z$_\odot$; see also][]{swinbank2004,nagao2012,rowlands2014,yang2017}. 

We can evaluate the systematic offset that a sub-solar metallicty (e.g. 0.5~Z$_\odot$) would induce on stellar mass. The metallicity dependence of $\alpha_\textrm{CO}$ adopted by \citet{tacconi2020,tacconi2018} is the geometrical average between the \citet{genzel2012} and \citet{bolatto2013} functions. At half-solar metallicity (corresponding to $12+\log(\textrm{O/H})\simeq8.4$ in the \citealt{pettini2004} scale), the conversion factor is 1.73 times larger than at Z$_\odot$, and so is the depletion timescale. Plugging this correction in Eq. \ref{eq:Mstar_no_lensing}, it turns out that the effect on stellar mass would be an increase by $\sim0.6$ dex at $z=1-3$, with the exact value of this shift slightly depending on redshift.



\section{Tables including results}

\tiny
\begin{longtable}{lcccccccc}
\caption{\label{tab:LprimeCO_all} Values of $L^\prime_{^{12}\textrm{CO}}$ of all detected transitions of the $z$-GAL sources.}\\
\hline
\hline
Source 	& redshift & \multicolumn{7}{c}{$\mu L^\prime_{^{12}\textrm{CO}}$ $[\rm 10^{10}\ K \, km \, s^{-1} pc^2]$}\\
	& & $(2-1)$ & $(3-2)$ & $(4-3)$ & $(5-4)$ & $(6-5)$ & $(7-6)$ & $(8-7)$  \\
\hline
\endfirsthead
\caption{continued.}\\
\hline
\hline
Source 	& redshift & \multicolumn{7}{c}{$\mu L^\prime_{^{12}\textrm{CO}}$ $[\rm 10^{10}\ K \, km \, s^{-1} pc^2]$}\\
	& & $(2-1)$ & $(3-2)$ & $(4-3)$ & $(5-4)$ & $(6-5)$ & $(7-6)$ & $(8-7)$  \\
\hline
\endhead
\hline
\endfoot
%
%
           HeLMS-1 &   1.9047 & 19.63$\pm$1.19 &  -- & 14.55$\pm$0.68 & 	 -- &  -- &  -- &  -- \\ 
           HeLMS-3 &   1.4199 & 31.35$\pm$2.40 & 25.69$\pm$1.92 &  -- & 	 -- &  -- &  -- &  -- \\ 
          HeLMS-11 &   2.4834 &  -- & 15.89$\pm$1.38 & 12.92$\pm$1.04 & 	 -- &  -- &  -- &  -- \\ 
          HeLMS-12 &   2.3699 &  -- & 33.31$\pm$2.33 & 31.19$\pm$2.01 & 	 -- &  -- &  -- &  -- \\ 
          HeLMS-14 &   1.6168 & 30.59$\pm$1.85 & 22.07$\pm$1.36 &  -- & 	 -- &  -- &  -- &  -- \\ 
          HeLMS-16 &   2.8187 &  -- & 22.17$\pm$3.17 &  -- & 	 -- &  -- &  -- &  -- \\ 
          HeLMS-17 E &   2.2983 &  -- &  -- &  1.20$\pm$0.27 & 	 -- &  0.94$\pm$0.23 &  -- &  -- \\ 
          HeLMS-17 W &   2.2972 &  -- &  -- &  5.18$\pm$0.40 & 	 -- &  5.02$\pm$0.28 &  -- &  -- \\
\end{longtable}

\begin{landscape}
\tiny
\begin{longtable}{lccccccccccccc}
\caption{\label{tab:Mgas_etc} Derived physical properties of the $z$-GAL sources. Dust modelling has been performed only for single sources or secure lensed multiples, therefore for all other cases several columns are not filled.}\\
\hline
\hline
Source 	& Lowest CO 	& $\mu L^\prime_{^{12}\textrm{CO}(1-0)}$ & $\mu M_{\textrm{mol},^{12}\textrm{CO}}$ & $\mu M_{\textrm{mol},850\mu\textrm{m}}$ & $\delta_{\textrm{GDR},^{12}\textrm{CO}}$ & $\mu L_\textrm{IR}$ & $\mu$SFR & $\tau_{\textrm{dep},^{12}\textrm{CO}}$ & $\mu^E M^\ast$ & $E$ & $\log(\mu\textrm{SFR}/\textrm{SFR}(MS))$\\
	& Transition & $[\rm 10^{10}\ K \, km \, s^{-1} pc^2]$ & $[10^{11}$ M$_\odot]$ & $[10^{11}$ M$_\odot]$ & & $[10^{11}$ L$_\odot]$ & $[$M$_\odot$ yr$^{-1}]$ & $[10^9$ yr$]$ & $[10^{11}$ M$_\odot]$ & & \\
\hline
\endfirsthead
\caption{continued.}\\
\hline
\hline
Source 	& Lowest CO 	& $\mu L^\prime_{^{12}\textrm{CO}(1-0)}$ & $\mu M_{\textrm{mol},^{12}\textrm{CO}}$ & $\mu M_{\textrm{mol},850\mu\textrm{m}}$ & $\delta_{\textrm{GDR},^{12}\textrm{CO}}$ & $\mu L_\textrm{IR}$ & $\mu$SFR & $\tau_{\textrm{dep},^{12}\textrm{CO}}$ & $\mu^E M^\ast$ & $E$ & $\log(\mu\textrm{SFR}/\textrm{SFR}(MS))$\\
	& Transition & $[\rm 10^{10}\ K \, km \, s^{-1} pc^2]$ & $[10^{11}$ M$_\odot]$ & $[10^{11}$ M$_\odot]$ & & $[10^{11}$ L$_\odot]$ & $[$M$_\odot$ yr$^{-1}]$ & $[10^9$ yr$]$ & $[10^{11}$ M$_\odot]$ & & \\
\hline
\endhead
\hline
\endfoot
           HeLMS-1 & $2-1$ &   23.09$\pm$1.40 &    9.24$\pm$0.56  	 &   13.26 $\pm$    0.89 	 &   63.32 $\pm$   18.60 	 &  467.39 $\pm$   32.02 	 &  5094.58 $\pm$   498.67 	 & 	   0.18 $\pm$    0.05 	 &     4.72 $\pm$     3.61 	 &  1.23 & 	    1.08 $\pm$    0.58 \\ 
           HeLMS-3 & $2-1$ &   36.89$\pm$2.82 &   14.75$\pm$1.13  	 &   12.16 $\pm$    1.37 	 &  105.36 $\pm$   18.63 	 &  387.38 $\pm$   37.81 	 &  4222.45 $\pm$   588.77 	 & 	   0.35 $\pm$    0.05 	 &    23.13 $\pm$    11.58 	 &  1.27 & 	    0.69 $\pm$    0.39 \\ 
          HeLMS-11 & $3-2$ &   24.07$\pm$2.10 &    9.63$\pm$0.84  	 &   20.84 $\pm$    1.58 	 &   70.54 $\pm$   13.01 	 &  318.67 $\pm$   14.77 	 &  3473.50 $\pm$   229.93 	 & 	   0.28 $\pm$    0.04 	 &     8.48 $\pm$     3.74 	 &  1.20 & 	    0.56 $\pm$    0.35 \\ 
          HeLMS-12 & $3-2$ &   50.47$\pm$3.54 &   20.19$\pm$1.42  	 &    9.98 $\pm$    1.29 	 &  271.50 $\pm$   78.34 	 &  337.84 $\pm$   18.07 	 &  3682.43 $\pm$   281.42 	 & 	   0.55 $\pm$    0.14 	 &    48.41 $\pm$    33.49 	 &  1.21 & 	    0.03 $\pm$    0.54 \\ 
          HeLMS-14 & $2-1$ &   35.99$\pm$2.17 &   14.40$\pm$0.87  	 &    7.36 $\pm$    1.33 	 &  146.60 $\pm$   30.45 	 &  195.88 $\pm$   14.12 	 &  2135.10 $\pm$   219.83 	 & 	   0.67 $\pm$    0.08 	 &    48.13 $\pm$    20.40 	 &  1.25 & 	    0.06 $\pm$    0.33 \\ 
          HeLMS-16 & $3-2$ &   33.60$\pm$4.81 &   13.44$\pm$1.92  	 &    7.42 $\pm$    1.23 	 &  220.62 $\pm$   50.15 	 &  425.34 $\pm$   19.46 	 &  4636.20 $\pm$   303.08 	 & 	   0.29 $\pm$    0.05 	 &    13.80 $\pm$     6.63 	 &  1.19 & 	    0.45 $\pm$    0.38 \\ 
          HeLMS-17 E & $4-3$ &    2.60$\pm$0.58 &    1.04$\pm$0.23  & -- & -- & -- & -- & -- & -- & -- & -- \\
          HeLMS-17 W & $4-3$ &   11.26$\pm$0.87 &    4.50$\pm$0.35  & -- & -- & -- & -- & -- & -- & -- & -- \\
\end{longtable}
\end{landscape}

\end{appendix}


\end{document}